\documentclass[10pt, journal]{IEEEtran}

\usepackage[utf8]{inputenc}
\usepackage{tikz}
\usepackage{rotating}
\usepackage{amsmath,amssymb,amsthm}
\usepackage{listings}
\usepackage{todonotes}
\usepackage{ wasysym }
\usepackage{pifont}
\usepackage{csvsimple}
\usepackage{siunitx}
\usepackage{tabularx}
\usepackage{xifthen}
\usepackage{subcaption}
\usepackage[T1]{fontenc}
\usepackage{dblfloatfix}
\usepackage{stmaryrd}
\usepackage{multirow}

\usepackage{footnote}

\usepackage{url}

\usepackage{tikz}
\usetikzlibrary{arrows}
\usetikzlibrary{positioning}  
\usetikzlibrary{shadows}
\usetikzlibrary{arrows,backgrounds,calc,chains,
	matrix,positioning,shapes,shapes.geometric,
	shapes.arrows, decorations.pathmorphing, decorations.pathreplacing}

\tikzstyle{vecArrow} = [thick, decoration={markings,mark=at position
	1 with {\arrow[semithick]{open triangle 60}}},
double distance=1.4pt, shorten >= 5.5pt,
preaction = {decorate},
postaction = {draw,line width=1.4pt, white,shorten >= 4.5pt}]

\newtheorem{definition}{Definition}

\newtheorem{example}{Example}

\usepackage{etoolbox}
\makeatletter
\patchcmd{\maketitle}{\@copyrightspace}{}{}{}
\makeatother

\usetikzlibrary{backgrounds,calc,chains,matrix,positioning,shapes,shapes.geometric,shapes.arrows}

\tikzset{%
	font={\footnotesize},
	vertex/.style={draw,circle,inner sep=0pt,minimum width=0.5cm,minimum height=0.5cm},
	zeroterm/.style={below,inner sep=0pt,font=\tiny}
}

\title{Advanced Simulation of Quantum Computations}

\author{Alwin Zulehner~\IEEEmembership{Student Member,~IEEE,} and Robert Wille~\IEEEmembership{Senior Member,~IEEE}\\
	alwin.zulehner@jku.at\hspace{8mm}robert.wille@jku.at}
\date{}

\newcommand{\ket}[1]{\ensuremath{\left|#1\right\rangle}} 
\usepackage[keeplastbox]{flushend}
\usepackage{qcircuit}

\begin{document}

\maketitle

\begin{abstract}
Quantum computation is a promising emerging technology which, compared to conventional computation, allows for substantial speed-ups e.g.~for integer factorization or database search. However, since physical realizations of quantum computers are in their infancy, a significant amount of research in this domain still relies on simulations of quantum computations on conventional machines. 
This causes a significant complexity which current state-of-the-art simulators try to tackle with a rather straight forward array-based representation and by applying massive hardware power.
There also exist solutions based on decision diagrams (i.e.~graph-based approaches) that try to tackle the exponential complexity by exploiting redundancies in quantum states and operations. However, these existing approaches do not fully exploit redundancies that are actually present.

In this work, we revisit the basics of quantum computation, investigate how corresponding quantum states and quantum operations can be represented even more compactly, and, eventually, simulated in a more efficient fashion.
This leads to a new graph-based simulation approach which outperforms state-of-the-art simulators (array-based as well as graph-based).
Experimental evaluations show that the proposed solution is capable of simulating quantum computations for more qubits than before, and in significantly less run-time (several magnitudes faster compared to previously proposed simulators). 
An implementation of the proposed simulator is publicly available online at \url{http://iic.jku.at/eda/research/quantum_simulation}.
\end{abstract}

\section{Introduction}
\label{sec:intro}

Quantum computation~\cite{NC:2000} has become a promising technology which has theoretically been proven to be superior to conventional computation for important applications. For example, quantum algorithms for integer factorization (Shor's algorithm~\cite{DBLP:journals/siamcomp/Shor97}) or database search (Grover's Search~\cite{DBLP:conf/stoc/Grover96}) have been proposed that lead to significant -- sometimes even exponential -- speedups compared to conventional computations. With respect to physical implementations, significant progress has been made in the recent years as well~\cite{hanneke2010realization,reed2012realization,debnath2016demonstration,monz2016realization,linke2017experimental}. 
The first publicly available quantum processor has been made accessible by IBM through their project \emph{IBM Q}~\cite{ibmQ}. Via IBM's cloud infrastructure, the community can access a quantum processor with 5 qubits (launched in March 2017) and 16 qubits (launched in June 2017), respectively, to conduct experiments. IBM further plans to increase the number of available qubits to 50 -- similar to Google's plans to provide a quantum chip with 49 qubits that demonstrates quantum supremacy~\cite{supremacy2017,gomes2018quantumcomputing}.

However, thus far, quantum computation remains an \linebreak emerging technology. This requires, besides others, that respective developments have to be conducted while still relying on conventional technologies. In particular, this is an issue when it comes to simulating quantum computations or corresponding quantum algorithms. Although these quantum computations describe approaches to solve several problems significantly faster than a conventional technology, they still have to be simulated on conventional machines thus far. Furthermore, simulation plays an important role in the verification of existing and future quantum computers. 

This causes a significant obstacle since basic and substantial concepts of quantum computations like superposition, entanglement, or measurement rely on exponentially large vector and matrix descriptions which additionally are composed of complex numbers. The majority of the existing methods for the simulation of quantum computations~\cite{DBLP:conf/pldi/GreenLRSV13,DBLP:journals/corr/WeckerS14,DBLP:journals/corr/SmelyanskiySA16,DBLP:conf/sc/HanerSST16,qxSimulator2017,pednault2017barrier,steiger2016projectq} address this problem using straight-forward methods like simple \mbox{1-dimensional} and 2-dimensional arrays, respectively.\footnote{Consequently, these methods are also called \emph{array-based simulators}.}
The resulting (exponential) complexity is then tackled by exploiting parallelism and applying massive hardware power such as supercomputers composed of thousands of nodes and more than a petabyte of distributed memory. But even then, quantum systems of rather limited size (today's practical limit is below 50 qubits~\cite{pednault2017barrier})
can be simulated -- additionally often requiring a significant amount of run-time (e.g.~up to several days). 
Also current roadmaps show that also future plans rely on the use of massive hardware power, e.g.~the authors of~\cite{DBLP:journals/corr/SmelyanskiySA16} expect to simulate 48-49 qubits on a machine with 4-10 petabytes of distributed memory.

In addition to that, simulators that utilize decision diagrams (e.g.~\cite{DBLP:books/daglib/0027785,DBLP:conf/esa/Samoladas08,DBLP:conf/rc/HiraishiI14}) 
have been proposed to tackle the exponential complexity of simulating quantum computations by exploiting redundancies. While decision diagrams have already been successfully used to solve exponential problems in the conventional domain (e.g.~in verification~\cite{MWBS:88} or synthesis~\cite{DSF:2004}) 
significantly faster than straight-forward solutions, simulators for quantum computations based on decision diagrams (i.e.~graph-based approaches) did not get established yet. 
In fact, the existing methods only exploit redundancies in a rather straight-forward fashion, which results in simulators that outperform \mbox{array-based} simulators for certain applications only.

In this work, we propose to use another type of decision diagram for the simulation of quantum computations that utilizes a decomposition scheme that is more natural to the occurring matrices and vectors -- allowing to exploit even more redundancies. 
To this end, we revisit the basics of quantum computations and investigate how corresponding matrices and vectors can represented in a more efficient fashion.
These endeavors eventually lead to a significantly more compact representation of quantum states and quantum operations than before, which exploits more redundancies in the corresponding description whenever possible. Furthermore, the natural decomposition scheme allows to realize dedicated manipulation algorithms efficiently -- leading to a new simulation method which clearly outperforms the current state-of-the-art.

In fact, the resulting compact representation allows for the simulation of well known quantum algorithms (such as Shor's Algorithm and Grover's Search) for more qubits than before.  
Finally, with respect to the run-time, a substantial drop can be observed: Instead of several days, the proposed approach is able to complete the simulations within hours -- in many cases even just minutes or seconds. These improvements can be obtained with respect to array-based simulators as well as graph-based simulators. 

This paper is structured as follows: Section~\ref{sec:background} revisits the basics of quantum computation. 
In Section~\ref{sec:simulation}, we investigate the obstacles of simulating quantum computations and review how the current state-of-the-art copes with these issues. Furthermore, the main idea of the utilized decision diagram is briefly sketched.
In Sections~\ref{sec:representation} and~\ref{sec:operations} we respectively discuss the resulting representation for vectors and matrices and the operations required to conduct the simulation in detail.
In Section~\ref{sec:discussion}, we discuss the limitations of the approach and analyze the complexity of the required operations.
Finally, the proposed solution is evaluated and compared to the \mbox{state-of-the-art} in Section~\ref{sec:exp}, while Section~\ref{sec:conclusion} concludes the paper. 

\section{Quantum Computation}
\label{sec:background}
Quantum computation significantly differs from the conventional computation paradigm. To make this work self-contained and to properly introduce our solution, we first briefly revisit the basics on how operations are conducted in this domain. While this ought to be sufficient to comprehend the remainder of this paper, we refer to~\cite{NC:2000} for a more detailed treatment.

\subsection{Quantum Bits}
\label{sec:qubits}

In conventional logic, information is represented by \emph{bits} which can be in one of two basis states 0 and 1. 
Similarly, quantum computations rely on so called \emph{quantum bits}~(\emph{qubits}) to represent internal states. Again, there exist two basis states, which -- using the Dirac notation -- are denoted~$\ket{0}$ and~$\ket{1}$.
However, in contrast to bits in conventional logic, qubits are not restricted to one of these basis states, but may additionally assume an (almost) arbitrary superposition (i.e.~a linear combination) of both.  More precisely, the state of a qubit is described by \mbox{$\ket{\psi} = \alpha_0 \cdot \ket{0} + \alpha_1 \cdot \ket{1}$}, 
where the complex factors $\alpha_0$ and $\alpha_1$ denote \emph{amplitudes} which indicate how much the qubit is related to the basis states.

The amplitudes of a quantum state~$\ket{\psi}$ must satisfy the normalization constraint \mbox{$\left|\alpha_0\right|^2 + \left|\alpha_1\right|^2 = 1$}.
While it is not possible to directly access the values of $\alpha_0$ and $\alpha_1$, it is possible to obtain one of the two basis states by measuring the qubit.
More precisely, the basis state $\ket{0}$ is obtained with probability $\left|\alpha_0\right|^2$, while $\ket{1}$ is obtained with probability~$\left|\alpha_1\right|^2$. The measurement collapses (i.e.~destroys) the superposition. 
The concepts discussed above can be generalized for quantum systems composed of multiple qubits. Since each qubit has exactly two basis states, a system composed of $n$ qubits has~$2^n$ basis states -- each one represented by $\ket{\{0,1\}^n}$. 
Overall, this accumulates in the following definition of a quantum state:

\begin{definition}
	Consider a quantum system composed of $n$ qubits. Then, all possible states of the system are of form
	\[\ket{\psi} = \sum_{x\in \{0,1\}^n} \alpha_x \cdot \ket{x}, \text{ where } \sum_{x\in \{0,1\}^n} \left|\alpha_x\right|^2 = 1 \text{ and } \alpha_x \in \mathbb{C}.\]
	The \emph{state} $\ket{\psi}$ can be also represented by a column vector $\psi = [\psi_i]$ with $0\le i < 2^n$ and $\psi_i = \alpha_x$, where $nat(x) = i$.
\end{definition}

Note that, to save space, vectors may be provided in their transposed form in the following  (indicated by~$\left[\cdot\right]^T$). That is, the single elements are listed horizontally rather than vertically. 
\begin{example}
	\label{ex:qubits}
	Consider a quantum system composed of two qubits which is in the state 
	\linebreak ${\ket{\psi} = \frac{1}{\sqrt{2}}\ket{00} + 0\cdot \ket{01} + 0\cdot \ket{10} +\frac{1}{\sqrt{2}}\ket{11}}$. This represents a valid state, since ${\left(\frac{1}{\sqrt{2}}\right)^2 + 0^2 + 0^2 + \left(\frac{1}{\sqrt{2}}\right)^2 = 1}$. 
	The corresponding state vector is 
	\[
	\psi = \left[ \frac{1}{\sqrt{2}}, 0, 0, \frac{1}{\sqrt{2}} \right]^T.
	\]
	Measuring this system yields one of the two basis states $\ket{00}$ or $\ket{11}$ -- both with probability of \mbox{$| \frac{1}{\sqrt{2}}|^2 = \frac{1}{2}$}. 
\end{example}

\subsection{Quantum Operations}
\label{sec:quantum_operations}

\emph{Quantum operations} are used to manipulate the current state of a quantum system. 
All of them except the measurement are thereby inherently reversible and can be represented by unitary matrices~$U$, i.e.~a complex square matrix whose
inverse is its conjugate transposed~\cite{NC:2000}.
The size of the matrix depends on the number of involved qubits. 
Important quantum operations for a single qubit are e.g.~
\[
X= \begin{bmatrix} 0 & 1 \\ 1 & 0 \end{bmatrix},
H= \frac{1}{\sqrt{2}}\begin{bmatrix} 1 & 1 \\ 1 & -1 \end{bmatrix}, \text{and }
Z= \begin{bmatrix} 1 & 0 \\ 0 & -1 \end{bmatrix}, 
\] 
where $X$ complements the current state of the qubit, $H$ sets the qubit into superposition, and~$Z$ changes the phase of the qubit, respectively.
An important operation involving two qubits is e.g.~
\[
CNOT= \begin{bmatrix} 
1 & 0 &0  & 0 \\ 
0 & 1 &0  & 0 \\
0 & 0 &  0 & 1 \\
0 & 0 &1  & 0  \\
\end{bmatrix}, \quad 
\]
which performs a so-called controlled inversion. Here, one qubit serves as \emph{control qubit}. The value the other qubit (i.e.~the \emph{target qubit}) is complemented if the \emph{control qubit} is in base state $\ket{1}$. Consequently, the resulting matrix is composed of the $2\times 2$ identity matrix in case that the \emph{control qubit} is in basis state $\ket{0}$ (left upper quadrant) and the single qubit matrix $X$ in case that the \emph{control qubit} is in basis state $\ket{1}$. This concept can  be easily extended to support single qubit gates that are controlled by multiple other qubits.\footnote{Note that the matrix grows exponentially with the number of control qubits.}

To evaluate a quantum operation with respect to a given quantum state, the corresponding matrix~$U$ has to be multiplied 
with the corresponding state vector~$\psi$. More precisely:
\begin{definition}\label{def:qua_op}
	Consider a quantum system composed of $n$ qubits with 
	\begin{itemize}
		\item a quantum operation~$U$ represented by a $2^n \times 2^n$ unitary matrix $U=\left[u_{i,j}\right]$ with $0 \le i,j < 2^n$ and
		\item a system state~$\ket{\psi}$ represented by a vector~$\psi=\left[\psi_i\right]$ with $0 \le i < 2^n$. 
	\end{itemize}
	Then, the output state~$\ket{\psi'}$ of the quantum system is defined by a vector $\psi' = U\cdot \psi$, i.e.~$\psi' = \left[\psi'_i\right]$ with
	\[
	\psi'_i = \sum_{k=0}^{2^n-1}u_{i,k}\cdot \psi_k, \quad \text{for $0 \le i < 2^n$.}
	\]  
\end{definition}

\begin{example}\label{ex:qua_op}
	Consider a quantum system composed of two qubits which is currently in state $\ket{\psi} = \ket{11}$. Applying a \emph{CNOT} operation yields
	\[
		\underbrace{\begin{bmatrix} 1 & 0 & 0 & 0 \\ 0 & 1 & 0 & 0 \\ 0 & 0 & 0 & 1 \\ 0 & 0 & 1 & 0 \end{bmatrix}}_{CNOT} \cdot \underbrace{\begin{bmatrix} 0 \\ 0 \\ 0 \\ 1 \end{bmatrix}}_{\psi} = \begin{bmatrix} 0+0+0+0 \\ 0+0+0+0 \\ 0+0+0+1 \\ 0+0+0+0 \end{bmatrix} = \begin{bmatrix} 0 \\ 0 \\ 1 \\ 0 \end{bmatrix} \equiv \ket{10}.
	\]

\end{example}

Quantum circuits are used as proper description for a sequence of quantum operations. A \emph{quantum circuit}~\cite{NC:2000} consists of a set of qubits, which are vertically aligned in a circuit diagram. The time axis is represented by a horizontal line for each qubit and read from left to right. Boxes on the time axis of a qubit indicate which quantum operation has to be applied. Note that measurement as reviewed in Section~\ref{sec:qubits} and illustrated in Example~\ref{ex:qubits} also counts as quantum operation in this context.

\begin{figure}
	
	\begin{center}\mbox{	
\Qcircuit @C=1.5em @R=1.5em {
	\lstick{\ket{q_0} = \ket{0}} & \gate{H} & \ctrl{1} & \meter & \qw \\
	\lstick{\ket{q_1} = \ket{0}} & \qw & \targ & \qw & \qw \\
}}
\end{center}
	
	\caption{Quantum circuit}
	\label{fig:quantum_circuit}
\end{figure}
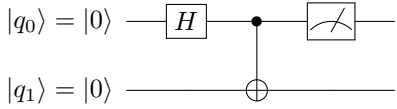

\begin{example}
	Consider the quantum circuit shown in \linebreak Fig.~\ref{fig:quantum_circuit}. The circuit contains two qubits, $q_0$ and $q_1$, which are both initialized with basis state $\ket{0}$. Consequently, the initial state is \mbox{$\ket{\psi} = \ket{00}$}. First, a Hadamard operation is applied to qubit~$q_0$, which is represented by a box labeled~\emph{H}. Then, a \emph{CNOT} operation is performed, where $q_0$ is the \emph{control qubit} (denoted by $\bullet$) and $q_1$ is the target qubit (denoted by~$\oplus$).
	Finally, qubit $q_0$ is measured (represented by the meter symbol), which collapses its superposition into one of the two basis states.
\end{example}

\section{Conducting Simulation}
\label{sec:simulation}

The basics reviewed in the previous section are sufficient to simulate the execution of quantum operations.
In fact, for a given sequence of quantum operations to be simulated, corresponding simulators simply have to conduct 
the multiplications of each operation matrix~$U$ with the respective intermediate quantum state~$\ket{\psi}$ as reviewed in 
Def.~\ref{def:qua_op} and illustrated in Example~\ref{ex:qua_op}. 
However, for actual quantum algorithms severe challenges emerge which significantly restrict 
today's capabilities to simulate quantum computations. 
In the following, these challenges are discussed -- followed by a summary of how state-of-the-art solutions currently deal with them.

\subsection{Exponential Growth}

A quantum circuit can be simulated by multiplying all matrices describing the quantum operation (from left to right) successively to the state vector. Therefore, all matrices have to be of dimension $2^n \times 2^n$. 
Since most quantum operations work on $k < n$ qubits only, their matrices have to be expanded to match the size of the state vector. 
To this end, an operation matrix for the remaining $n-k$ qubits is required. Since they shall not be affected by the gate, a $2 \times 2$ identity matrix~$I_2$ is used for this purpose. The overall $2^n \times 2^n$-matrix is eventually obtained by forming the Kronecker product of all these matrices.

\begin{example}
	\label{ex:simulation}
	Consider again the quantum circuit shown in Fig.~\ref{fig:quantum_circuit} with state $\ket{q_0q_1}=\ket{00}$ as input. The first operation of the circuit is a Hadamard operation, which is applied to qubit $q_0$. Since this operation shall not affect $q_1$, we form the Kronecker product of \emph{H} and the identity matrix $I_2$, i.e.
	\[
	H \otimes I_2 = \frac{1}{\sqrt{2}}\begin{bmatrix} 1 & 1 \\ 1 & -1 \end{bmatrix}\otimes \begin{bmatrix}1 & 0 \\ 0 & 1\end{bmatrix} = \frac{1}{\sqrt{2}} \begin{bmatrix}1 & 0&  1&  0 \\ 0 &  1& 0 & 1 \\ 1 & 0 & -1 & 0 \\ 0 & 1 & 0 & -1\end{bmatrix}.
	\] 
	Multiplying this matrix with the state vector yields
	\[
	\psi' = \frac{1}{\sqrt{2}} \begin{bmatrix}1 & 0&  1&  0 \\ 0 &  1& 0 & 1 \\ 1 & 0 & -1 & 0 \\ 0 & 1 & 0 & -1\end{bmatrix}\cdot\begin{bmatrix}1 \\ 0\\0\\0\end{bmatrix} = \frac{1}{\sqrt{2}}\begin{bmatrix}1\\0\\1\\0\end{bmatrix}.
	\]
	Applying the \emph{CNOT} operation yields 
	\[
	\psi'' = \begin{bmatrix}1 & 0&  0&  0 \\ 0 & 1 & 0 & 0 \\ 0 & 0 & 0 & 1 \\ 0 &0 & 1 & 0\end{bmatrix}\cdot \frac{1}{\sqrt{2}}\begin{bmatrix}1 \\ 0\\1\\0\end{bmatrix} = \frac{1}{\sqrt{2}}\begin{bmatrix}1\\0\\0\\1\end{bmatrix}.
	\]
\end{example}

Since both, the state vectors as well as the operation matrices grow exponentially with respect to the number~$n$ of qubits, a crucial obstacle becomes evident: 
The simulation of quantum computations requires an exponential amount of space. The same complexity applies for the measurement of a quantum state, since, because of superposition, also  
the state vector may contain an exponentially large number of non-zero entries.

Now, one might think that a local consideration of qubits during the simulation avoids this exponential blow-up: Instead of forming a 
$2^n \times 2^n$-matrix using the Kronecker product, a simple application of an operation matrix to only those qubits which are actually affected might be sufficient.
Unfortunately, this is not possible, since \emph{entanglement}, which is one of the main concepts that make quantum computations superior to conventional computations, frequently occurs~\cite{NC:2000}. Two qubits are entangled if their state cannot be described without the other. An example illustrates the concept:

\addtocounter{example}{-1}
\begin{example}[continued]
	\label{ex:entanglement}
	Consider again the quantum state $\ket{\psi''}$ from above.
	If we e.g.~measure $q_0$, this qubit collapses to the basis state $\ket{0}$ or $\ket{1}$ with a probability \mbox{$\left|\frac{1}{\sqrt{2}}\right|^2 = \frac{1}{2}$}. 
	But due to the nature of~$\ket{\psi''}$, 
	this measurement also affects~$q_1$. More precisely, if the measurement yields e.g.~the basis state~$\ket{0}$ for~$q_0$, the new state vector is $\psi' = \left[1, 0, 0, 0\right]^T$, i.e.~also~$q_1$ collapses to the basis state~$\ket{0}$ (although not explicitly measured). 
	This happens because~$\ket{\psi''}$ represents a state in which both qubits are entangled.
\end{example}

Since actual quantum computations frequently use entangled states, a local consideration of qubits affected by an operation is often not possible. Instead the complete (exponential) state vector is required.

\subsection{State-of-the-Art Solutions}
\label{sec:sota}

In the recent past, researchers and engineers intensely considered this problem and developed corresponding solutions for the simulation of quantum computations. 
Most of them are so-called \emph{array-based approaches}, which are, however,
rather limited, since they rely on a straight-forward representation of quantum states and operations (besides minor optimization, mainly representations such as simple 1-dimensional and \mbox{2-dimensional} arrays are employed).
Consequently, only experimental results for quantum systems with up to 34 qubits were reported on Desktop machines. 
In order to simulate quantum systems composed of more qubits, solutions exploiting massive hardware power (supercomputers composed of thousands of nodes and more than a petabyte of distributed memory) are applied. But even then, quantum systems with less than 50 qubits are today's practical limit~\cite{DBLP:journals/corr/SmelyanskiySA16,pednault2017barrier}.

Besides that, solutions based on decision diagrams (so-called~\emph{graph-based simulators}~\cite{DBLP:books/daglib/0027785,DBLP:conf/esa/Samoladas08,DBLP:conf/rc/HiraishiI14}) have been proposed by researchers that try to exploit redundancies to gain a more compact representation of state vectors and matrices. One such type of decision diagrams suited for representing quantum computation are \emph{Quantum Information Decision Diagrams} (QuIDDs~\cite{viamontes2003improving}), which -- like \emph{Binary Decision Diagrams} (i.e.~BDDs~\cite{Bry:86}) in conventional design -- aim for a rather compact  representation in many cases.

Overall, the current state-of-the-art approaches (\emph{array-based} as well as \emph{graph-based} ones) can be summarized as follows:

\begin{itemize}

	\item \emph{LIQ\textit{Ui}\ket{}}~\cite{DBLP:journals/corr/WeckerS14}: Microsoft's tool suite for quantum computation with an integrated simulator which relies on a straight-forward representation and, thus, also can simulate systems with up to approximately 30 qubits only, when used on a Desktop machine with 32\,GB RAM (still requires substantial run-times of up to several days). 
	
	\item \emph{qHiPSTER}~\cite{DBLP:journals/corr/SmelyanskiySA16}: A quantum high performance software testing environment developed in Intel's Parallel Computing Lab. Here, parallel algorithms are utilized which are executed on 1000 compute nodes with 32\,TB RAM distributed across these nodes.
	Even with this massive hardware power, quantum systems of rather limited size (not more than 40 qubits) can be simulated.
	
	\item \emph{Quantum Emulator proposed in}~\cite{DBLP:conf/sc/HanerSST16}: A solution which utilizes a higher level description of quantum computations (e.g.~addition, quantum Fourier transformation, etc.) to directly compute intermediate results instead of applying the individual quantum operations successively. Experimental results are provided for systems with up to 36 qubits which, again, were accomplished with massive hardware power, i.e.~a supercomputer similar to the one used for \emph{qHiPSTER}.

	\item  \emph{QX}~\cite{qxSimulator2017}: A high-performance array-based quantum computer simulation platform developed in the QuTech Computer Engineering Lab at Delft University. 
	The simulator tries to parallelize the application of quantum gates to improve the performance. The authors state that \emph{QX} allows for simulation of 34 fully entangled qubits on a single node using 270 GB of memory.
	
	\item \emph{ProjectQ}~\cite{steiger2016projectq}: ProjectQ is a software framework for quantum computing that started at the ETH Zurich. The contained high-performance array-based simulator allows to simulate up to approximately 30 qubits on a desktop machine. ProjectQ additionally contains an emulator, which can determine e.g.~the result for Shor's algorithm significantly faster than the simulator by employing conventional shortcuts (e.g.~arithmetic components are computed conventionally instead of using a quantum circuit).

	\item \emph{QuIDDPro}~\cite{viamontes2004high,DBLP:books/daglib/0027785} is a graph-based simulator based on QuIDDs, which allows e.g.~to simulate Grover's algorithm significantly faster than with array-based solutions by exploiting certain redundancies in the occurring state vectors and unitary matrices. However, the performance for simulating computations such as \emph{Quantum Fourier Transformation} or \emph{Shor's Algorithms} is rather limited since QuIDDs require a non-scalable number of decision diagram nodes for these cases (cf.~Section~\ref{sec:exp}).
	
\end{itemize}

\subsection{General Idea}

In this work, we investigate how quantum states and quantum operations can be represented more compactly so that an efficient simulation becomes possible.
To this end, we utilize a decomposition scheme that is more natural to state vectors and unitary matrices used in simulation of quantum computations -- allowing for a more compact representation and for developing efficient manipulation algorithms. In fact, some of these operations often boil down to rearranging pointers in the decision diagram.
The general idea is motivated by the decomposition scheme of \emph{Quantum Multiple-Valued Decision Diagrams} (QMDDs~\cite{DBLP:journals/tcad/NiemannWMTD16}), which have not yet been utilized in the context of simulating quantum computations. They rather have been applied to efficiently solve design tasks such as verification~\cite{WGMD:2009,DBLP:conf/rc/NiemannWD14} and synthesis~\cite{DBLP:conf/aspdac/NiemannWD14,niemann2016logic,niemann2018cliffordt,zulehner2017one}.
In the following sections, we introduce and discuss the proposed representations as well as the required manipulation algorithms in detail.

\section{Representations for Quantum Simulation}
\label{sec:representation}

In this section, we discuss the proposed representation for state vectors and unitary matrices required for quantum simulation.
To this end, we describe a compact representation for state vectors which, afterwards, is extended by a second dimension -- leading to a compact representation 
for quantum operations. 

\subsection{Representation of State Vectors}
\label{sec:vector}

As discussed in Section~\ref{sec:qubits}, a system composed of $n$ qubits is represented by a state vector of size $2^n$ -- an exponential representation.
However, a closer look at state vectors unveils that they are frequently composed of redundant entries which provide ground for a more compact representation.

\begin{example}\label{ex:dd_vector}
	Consider a quantum system with $n=3$ qubits situated in a state given by the following vector:
	\[
	 \psi = \left[0, 0, \frac{1}{2}, 0, \frac{1}{2}, 0, -\frac{1}{\sqrt{2}}, 0\right]^T.
	\]	
	Although of exponential size ($2^3=8$ entries), this vector only assumes three different values, namely $0$, $\frac{1}{2}$, and $-\frac{1}{\sqrt{2}}$.
\end{example}

This redundancy can be exploited for a more compact representation. To this end, decision diagram techniques are employed. For conventional computations, e.g.~\emph{Binary Decision Diagrams} (BDDs, \cite{Bry:86}) are very well known. Here, a decomposition scheme is employed which reduces a function to be represented into corresponding sub-functions. Since they also usually include redundancies, equivalent sub-functions result which can be shared -- eventually yielding a much more compact representation. 
In a similar fashion, the concept of decomposition can also be applied to represent state vectors in a more compact fashion.

More precisely, similar to decomposing a function into \mbox{sub-functions}, we decompose a given state vector with its complex entries into sub-vectors. To this end, consider a quantum system with qubits $q_0, q_1, \ldots q_{n-1}$, whereby~$q_0$ represents the most significant qubit.\footnote{Note that, with the terminology \emph{most-significant qubit}, we refer to a position in the basis states of a quantum system, rather than to the importance of the qubit itself.} 
Then, the first $2^{n-1}$ entries of the corresponding state vector represent the amplitudes for the basis states with~$q_0$ set to $\ket{0}$; the other entries represent the amplitudes for states with $q_0$ set to~$\ket{1}$. This decomposition is represented in a decision diagram structure by a node labeled~$q_0$ and two successors leading to nodes representing the sub-vectors. The sub-vectors are recursively decomposed further until vectors of size~1 (i.e.~a complex number) results. This eventually represents the amplitude~$\alpha_i$ for the complete basis state and is given by a terminal node.
During these decompositions, equivalent sub-vectors can be represented by the same nodes -- allowing for sharing and, hence a reduction of the complexity of the representation.
An example illustrates the idea.

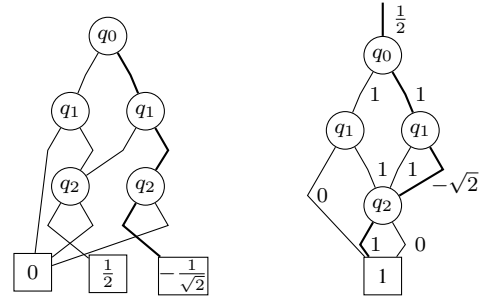
\begin{figure}
	\centering
	\begin{subfigure}[b]{0.2\textwidth}
		\centering
		\begin{tikzpicture}[terminal/.style={draw,rectangle,inner sep=0pt}]
		\matrix[matrix of nodes,ampersand replacement=\&,every node/.style={vertex},column sep={1cm,between origins},row sep={1cm,between origins}] (qmdd) {
			\& \node (n1) {$q_0$}; \& \\
			\node[xshift=0.5cm] (n2a) {$q_1$}; \& \node[xshift=0.5cm] (n2b) {$q_1$}; \& \\
			\node[xshift=0.5cm] (n3a) {$q_2$}; \& \node[xshift=0.5cm] (n3b) {$q_2$}; \& \\[0.2cm]
			\node[terminal] (t0){$0$}; \& \node[terminal] (t1) {$\frac{1}{2}$}; \& \node[terminal] (t2) {$-\frac{1}{\sqrt{2}}$};\\
		};
		
		\draw (n1) -- ++(240:0.6cm) -- (n2a);
		\draw[thick] (n1) -- ++(300:0.6cm) -- (n2b);
		
		\draw (n2a) -- ++(240:0.6cm) -- (t0);
		\draw (n2a) -- ++(300:0.6cm) -- (n3a);

		\draw (n2b) -- ++(240:0.6cm) -- (n3a);
		\draw[thick] (n2b) -- ++(300:0.6cm) -- (n3b);
		
		\draw (n3a) -- ++(240:0.6cm) -- (t1);
		\draw (n3a) -- ++(300:0.6cm) -- (t0);
		
		\draw[thick] (n3b) -- ++(240:0.6cm) -- (t2);
		\draw (n3b) -- ++(300:0.6cm) -- (t0);	
		\end{tikzpicture}
		\caption{Without edge weights}
		\label{fig:dd_vector_a}		
		\end{subfigure}
		\begin{subfigure}[b]{0.2\textwidth}
			\centering
		\begin{tikzpicture}[terminal/.style={draw,rectangle,inner sep=0pt}]	
		\matrix[matrix of nodes,ampersand replacement=\&,every node/.style={vertex},column sep={1cm,between origins},row sep={1cm,between origins}] (qmdd2) {
			\& \node (m1) {$q_0$}; \& \\
			\node[xshift=0.5cm] (m2a) {$q_1$}; \& \node[xshift=0.5cm] (m2b) {$q_1$}; \& \\
			\& \node (m3) {$q_2$}; \& \\
			\& \node[terminal] (t3) {1}; \& \\
		};
		
		\draw[thick] ($(m1)+(0,0.7cm)$) -- (m1) node[right, midway]{$\frac{1}{2}$};
		
		\draw (m1) -- ++(240:0.6cm) -- (m2a) node[right, at start] {$1$};
		\draw[thick] (m1) -- ++(300:0.6cm) -- (m2b) node[right, at start] {\bf $1$};
		
		\draw (m2a) -- ++(240:1.0cm) -- (t3) node[right, at start] {$0$};;
		\draw (m2a) -- ++(300:0.6cm) -- (m3) node[right, at start] {$1$};

		\draw (m2b) -- ++(240:0.6cm) -- (m3) node[right, at start] {$1$};
		\draw[thick] (m2b) -- ++(300:0.6cm) -- (m3) node[right, midway] {\bf $-\sqrt{2}$};
		
		\draw[thick] (m3) -- ++(240:0.6cm) -- (t3) node[right, at start] {\bf $1$};
		\draw (m3) -- ++(300:0.6cm) -- (t3) node[right, at start] {$0$};
		
		\end{tikzpicture}
		\caption{With edge weights}
		\label{fig:dd_vector_b}		
		\end{subfigure}
	\caption{Representation of the state vector}
	\label{fig:dd_vector}
\end{figure}

\begin{example}
	\label{ex:dd}
	Consider again the quantum state from Example~\ref{ex:dd_vector}.
	Applying the decompositions described above yields a decision diagram as shown in Fig.~\ref{fig:dd_vector_a}.	
	The left (right) outgoing edge of each node labeled~$q_i$ points to a node representing the sub-vector  
	with all amplitudes for the basis states with~$q_i$ set to $\ket{0}$ ($\ket{1}$).
	Following a path from the root to the terminal yields the respective entry. For example,
	following the path highlighted bold in Fig.~\ref{fig:dd_vector_a} provides the amplitude
	for the basis state with $q_0=\ket{1}$ (right edge), $q_1=\ket{1}$ (right edge), and $q_2=\ket{0}$ (left edge), i.e.~$-\frac{1}{\sqrt{2}}$ which is exactly the amplitude for basis state $\ket{110}$ (seventh entry in the vector from Example~\ref{ex:dd_vector}).
	Since some sub-vectors are equal (e.g.~$\left[\frac{1}{2},0\right]^T$	
	represented by the left node labeled~$q_2$), sharing is possible. 	
\end{example}

However, even more sharing is possible. In fact, many entries of the state vectors differ in a common factor only (e.g.~the state vector from Example~\ref{ex:dd_vector} has entries~$\frac{1}{2}$ and~$-\frac{1}{\sqrt{2}}$ which differ by the factor~$-\sqrt{2}$ only).
This is additionally exploited in the proposed representation by denoting common factors of amplitudes as weights to the edges of the
decision diagram. Then, the value of an amplitude for a basis state is determined by not only following the path from the root to the terminal, but
additionally multiplying the weights of the edges along this path. Again, an example illustrates the idea.

\begin{example}
	
	Consider again the quantum state from Example~\ref{ex:dd_vector} and the corresponding decision diagram shown in Fig.~\ref{fig:dd_vector_a}.	
	As can be seen, the sub-trees rooting the node labeled~$q_2$ are structurally equivalent and only differ in their terminal values. Moreover, they
	represent sub-vectors~$\left[\frac{1}{2},0\right]^T$ and~$\left[-\frac{1}{\sqrt{2}},0\right]^T$
	which only differ in a common factor.
	
	In the decision diagram shown in Fig~\ref{fig:dd_vector_b}, both sub-trees are merged. This is possible since the corresponding value of the amplitudes is now determined not by the terminals, but the weights on the respective paths. As an example, consider again the path highlighted bold representing the	
	amplitude for the basis state~$\ket{110}$. Since this path includes the weights $\frac{1}{2}$, 1, $-\sqrt{2}$, and $1$,
	an amplitude value of $\frac{1}{2}\cdot 1 \cdot (-\sqrt{2}) \cdot 1 = -\frac{1}{\sqrt{2}}$ results.

\end{example}

Note that, of course, various possibilities exist to factorize an amplitude. Hence, we apply a normalization which assumes the left edge to inherit a weight of~$1$. More precisely, the weights $w_l$ and $w_r$ of the left and right edge are both divided by $w_l$ and this common factor is propagated upwards to the parents of the node. If $w_l =0$, the node is normalized by propagating $w_r$ upwards to the parents of the node.\footnote{Applying a fixed normalization scheme, a representation which is even canonic (w.r.t~qubit order) results. However, since canonicity is not further relevant for the purpose of simulation, this issue is not discussed in detail in this work.} 

The resulting representation discussed above leads to the following definition. 

\begin{definition}
	\label{def:dd_sv}
The resulting decision diagram for representing a $2^n$-dimensional state vector is a directed acyclic graph with one terminal node labeled 1 that has no successors and represents a 1-dimensional vector with the element 1. All other nodes are labeled $q_i$, $0\le i < n$ (representing a partition over qubit $q_i$) and have two successors. Additionally, there is an edge pointing to the root node of the decision diagram. This edge is called \emph{root edge}. Each edge of the graph has attached a complex number as weight. An entry of the state vector is then determined by the product of all edge weights along the path from the root towards the terminal. 
Without loss of generality, the nodes of the decision diagram are ordered by the significance of their label, i.e.~the successor of a node labeled $q_i$ are labeled with a less significant qubit $q_j$.
The decision diagram is reduced, i.e.~nodes where both outgoing edges point to the same successor and have attached the same weight (i.e.~both sub-vectors are equal) are removed. Finally, the nodes are normalized, which means that all edges-weights are divided by the first non-zero weight. The common factor is propagated upwards in the decision diagram.

\end{definition}

\subsection{Representation of Matrices}
\label{sec:matrices}

As discussed in Section~\ref{sec:quantum_operations}, quantum operations are described by unitary 
matrices. 
Similar to state vectors, these matrices include redundancies, 
which can be represented in a more compact fashion. 
To this end, we extend the proposed decomposition scheme for state vectors by a second dimension -- yielding a decomposition scheme for $2^n\times 2^n$ matrices. 

The entries of a unitary matrix $U = [u_{i,j}]$ 
indicate how much the operation~$U$ affects the mapping from a basis state~$\ket{i}$ to a basis state~$\ket{j}$.
Considering again a quantum system with qubits $q_0, q_1, \ldots q_{n-1}$, whereby w.l.o.g.~$q_0$ represents the most significant qubit,
the matrix~$U$ is decomposed into four sub-matrices with dimension $2^{n-1}\times 2^{n-1}$: 
All entries in the left upper sub-matrix (right lower sub-matrix) provide the values describing the mapping from basis states $\ket{i}$ to $\ket{j}$ with both
assuming $q_0 =\ket{0}$ ($q_0 =\ket{1}$). All entries in the right upper sub-matrix (left lower sub-matrix) provide the values describing the mapping from basis states $\ket{i}$ with $q_0 =\ket{1}$ to $\ket{j}$ with $q_0 =\ket{0}$ ($q_0 =\ket{0}$ to $q_0 =\ket{1}$). 
This decomposition is represented in a decision diagram structure by a node labeled~$q_0$ and four successors leading to nodes representing the sub-matrices. The sub-matrices are recursively decomposed further until a $1\times 1$ matrix~(i.e.~a complex number) results. This eventually represents the value~$u_{i,j}$ for the corresponding mapping. Also during these decompositions, equivalent sub-matrices are represented by the same nodes and weights  as well as a corresponding normalization scheme (as applied for the representation of state vectors) is employed.
Note that for a simpler graphical notation, we use zero stubs to indicate zero matrices (i.e.~matrices that contain zeros only) and omit edge weights that are equal to one.
Again, an example illustrates the idea.

\begin{figure}
	\begin{subfigure}[b]{0.15\textwidth}
		\centering
		\begin{tikzpicture}[terminal/.style={draw,rectangle,inner sep=0pt}]
		\matrix[matrix of nodes,ampersand replacement=\&,every node/.style={vertex},column sep={1cm,between origins},row sep={1.2cm,between origins}] (qmdd) {
			\node[draw = none] (top) {}; \\
			\node (n1) {$q_0$}; \\
			\node[terminal] (t){1};\\
		};
		
		\draw (n1) -- ++(240:0.6cm) -- (t);
		\draw (n1) -- ++(260:0.6cm) -- (t);
		\draw (n1) -- ++(280:0.6cm) -- (t);
		\draw (n1) -- ++(300:0.6cm) node[right,midway] {$-1$} -- (t);
		
		\draw (top) -- (n1) node[right,midway] {$\frac{1}{\sqrt{2}}$};
		\end{tikzpicture}
		\caption{$H$}
		\label{fig:dd_h}		
	\end{subfigure}
	\begin{subfigure}[b]{0.15\textwidth}
		\centering
		\begin{tikzpicture}[terminal/.style={draw,rectangle,inner sep=0pt}]
		\matrix[matrix of nodes,ampersand replacement=\&,every node/.style={vertex},column sep={1cm,between origins},row sep={1.2cm,between origins}] (qmdd) {
			\node[draw = none] (top) {}; \\
			\node (n1) {$q_1$}; \\
			\node[terminal] (t){1};\\
		};
		
		\draw (n1) -- ++(240:0.6cm) -- (t);
		\draw (n1) -- ++(260:0.4cm) node[zeroterm]{0};
		\draw (n1) -- ++(280:0.4cm) node[zeroterm]{0};
		\draw (n1) -- ++(300:0.6cm) -- (t);
		
		\draw (top) -- (n1) node[right,midway] {};
		\end{tikzpicture}
		\caption{$I_2$}
		\label{fig:dd_i}		
	\end{subfigure}
	\begin{subfigure}[b]{0.15\textwidth}
		\centering
		\begin{tikzpicture}[terminal/.style={draw,rectangle,inner sep=0pt}]
		\matrix[matrix of nodes,ampersand replacement=\&,every node/.style={vertex},column sep={1cm,between origins},row sep={1cm,between origins}] (qmdd) {
			\node (n1) {$q_0$}; \\
			\node (n2) {$q_1$}; \\
			\node[terminal] (t){1};\\
		};
		
		\draw (n1) -- ++(240:0.6cm) -- (n2);
		\draw (n1) -- ++(260:0.6cm) -- (n2);
		\draw[thick] (n1) -- ++(280:0.6cm) -- (n2);
		\draw (n1) -- ++(300:0.6cm) node[right,midway] {$-1$} -- (n2);

		\draw[thick] (n2) -- ++(240:0.6cm) -- (t);
		\draw (n2) -- ++(260:0.4cm) node[zeroterm]{0};
		\draw (n2) -- ++(280:0.4cm) node[zeroterm]{0};
		\draw (n2) -- ++(300:0.6cm) -- (t);
		
		\draw[thick] ($(n1)+(0,0.7cm)$) -- (n1) node[right, midway]{$\frac{1}{\sqrt{2}}$};
		\end{tikzpicture}
		\caption{$U=H\otimes I_2$}
		\label{fig:dd_kronecker}		
	\end{subfigure}
	\caption{Representation of matrices}
	\label{fig:dd_matrices}
\end{figure}
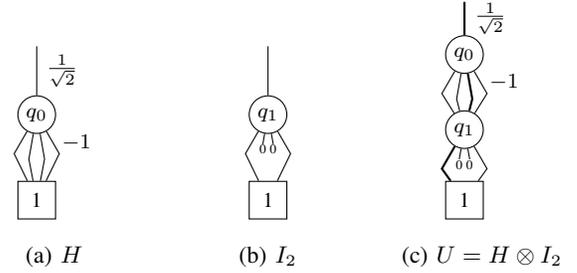

\begin{example}
	\label{ex:kronecker}
	Consider again the matrices of $H$, $I_2$, and $U = H \otimes I_2$ from Example~\ref{ex:simulation}. Fig.~\ref{fig:dd_matrices} shows the corresponding decision diagram representations. 
	Following the path highlighted  bold in Fig.~\ref{fig:dd_kronecker} defines the entry~$u_{0,2}$: a mapping from $\ket{0}$ to $\ket{1}$ for $q_0$ (third edge from the left) and from $\ket{0}$ to $\ket{0}$ for $q_1$ (first edge). Consequently the path describes the entry for a mapping from $\ket{00}$ to $\ket{10}$.  
	Multiplying all factors on the path (including the \emph{root edge}) yields $\frac{1}{\sqrt{2}}\cdot 1\cdot 1 = \frac{1}{\sqrt{2}}$, which is the value of~$u_{0,2}$.
\end{example}

The concepts described above yield to the definition of a decision diagram representing a unitary matrix as follows.
\begin{definition}
\label{def:dd_matrix}
The resulting decision diagram for representing a $2^n\times 2^n$-dimensional unitary matrix is a directed acyclic graph with one terminal node labeled 1 that has no successors and represents a $1\times 1$ matrix with the element 1. All other nodes are labeled $q_i$, $0\le i < n$ (representing a partition over qubit $q_i$) and have two successors. Additionally, there is an edge pointing to the root node of the decision diagram. This edge is called \emph{root edge}. Each edge of the graph has attached a complex number as weight. An entry of the unitary matrix is then determined by the product of all edge weights along the path from the root towards the terminal. Without loss of generality, the nodes of the decision diagram are ordered by the significance of their label, i.e.~the successor of a node labeled $q_i$ are labeled with a less significant qubits $q_j$.
The decision diagram is reduced, i.e.~nodes where all outgoing edges point to the same successor and have attached the same weight (i.e.~all four sub-matrices are equal) are removed. Finally, the nodes are normalized, which means that all edges-weights are divided by the first non-zero weight. The common factor is propagated upwards in the decision diagram.
	
\end{definition}

\section{Conducting Quantum Simulations}
\label{sec:operations}

With the availability of a compact representation for state vectors and unitary matrices, it is left to provide corresponding methods 
for conducting quantum operations, i.e.~forming the Kronecker product, multiplying vectors with matrices, as well as measuring the quantum system. 
Since the applied decomposition scheme is natural to vectors and matrices, we can efficiently implement these required operations.

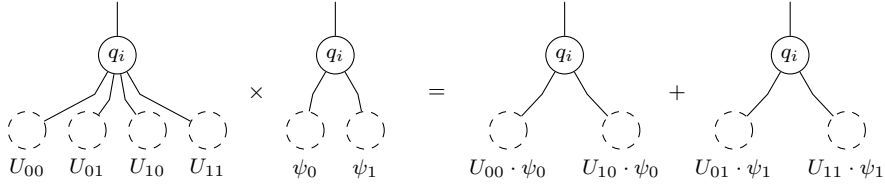
\begin{figure*}
	\centering
	\begin{tikzpicture}[terminal/.style={draw,rectangle,inner sep=2pt}]
	\matrix[matrix of nodes,ampersand replacement=\&,every node/.style={vertex},column sep={0.8cm,between origins},row sep={1cm,between origins}] (qmdd) {
		\& \& \node (top)[draw = none] {};\& \& \\
		\& \& \node (n1) {$q_{i}$};\& \& \\
		\node[dashed, xshift=0.4cm] (n2) {$\phantom{q_i}$}; \& \node[dashed, xshift=0.4cm] (n3) {$\phantom{q_i}$}; \& \node[dashed, xshift=0.4cm] (n4) {$\phantom{q_i}$}; \& \node[dashed, xshift=0.4cm] (n5) {$\phantom{q_i}$}; \& \\
	};
	
	\draw (top) -- (n1);
	
	\draw (n1) -- ++(240:0.6cm) -- (n2);
	\draw (n1) -- ++(260:0.6cm) -- (n3);
	\draw (n1) -- ++(280:0.6cm) -- (n4);
	\draw (n1) -- ++(300:0.6cm) -- (n5);
	
	\matrix[matrix of nodes, right=0.25cm of qmdd, ampersand replacement=\&,every node/.style={vertex},column sep={0.8cm,between origins},row sep={1cm,between origins}] (qmdd2) {
		\& \node (top2)[draw = none] {};\& \\
		\& \node (m1) {$q_{i}$};\& \\
		\node[dashed, xshift=0.4cm] (m2) {$\phantom{q_i}$}; \& \node[dashed, xshift=0.4cm] (m3) {$\phantom{q_i}$}; \& \\
	};
	
	\draw (top2) -- (m1);
	
	\draw (m1) -- ++(240:0.6cm) -- (m2);
	\draw (m1) -- ++(300:0.6cm) -- (m3);				
	
	\matrix[matrix of nodes, right=0.5cm of qmdd2, ampersand replacement=\&,every node/.style={vertex},column sep={1.5cm,between origins},row sep={1cm,between origins}] (qmdd3) {
		\& \node (top3)[draw = none] {};\& \\
		\& \node (o1) {$q_{i}$};\& \\
		\node[dashed, xshift=0.75cm] (o2) {$\phantom{q_i}$}; \& \node[dashed, xshift=0.75cm] (o3) {$\phantom{q_i}$}; \& \\
	};
	
	\draw (top3) -- (o1);
	
	\draw (o1) -- ++(240:0.6cm) -- (o2);
	\draw (o1) -- ++(300:0.6cm) -- (o3);
	
	\matrix[matrix of nodes, right=-0.25cm of qmdd3, ampersand replacement=\&,every node/.style={vertex},column sep={1.5cm,between origins},row sep={1cm,between origins}] (qmdd4) {
		\& \node (top4)[draw = none] {};\& \\
		\& \node (l1) {$q_{i}$};\& \\
		\node[dashed, xshift=0.75cm] (l2) {$\phantom{q_i}$}; \& \node[dashed, xshift=0.75cm] (l3) {$\phantom{q_i}$}; \& \\
	};
	
	\draw (top4) -- (l1);
	
	\draw (l1) -- ++(240:0.6cm) -- (l2);
	\draw (l1) -- ++(300:0.6cm) -- (l3);

	\draw (n2.south) node[anchor=north] {$U_{00}$};
	\draw (n3.south) node[anchor=north] {$U_{01}$};
	\draw (n4.south) node[anchor=north] {$U_{10}$};
	\draw (n5.south) node[anchor=north] {$U_{11}$};
	
	\draw (m2.south) node[anchor=north] {$\psi_{0}$};
	\draw (m3.south) node[anchor=north] {$\psi_{1}$};
	
	\draw (o2.south) node[anchor=north] {$U_{00}\cdot \psi_{0}$};
	\draw (o3.south) node[anchor=north] {$U_{10}\cdot \psi_{0}$};
	
	\draw (l2.south) node[anchor=north] {$U_{01}\cdot \psi_{1}$};
	\draw (l3.south) node[anchor=north] {$U_{11}\cdot \psi_{1}$};
	
	\draw ($(n5)!0.5!(m2) + (0,0.5)$) node {$\times$};
	
	\draw ($(o3)!0.5!(l2) + (0,0.5)$) node {$+$};
	
	\draw ($(m3)!0.5!(o2) + (0,0.5)$) node {$=$};

	\end{tikzpicture}	
	
	\caption{Multiplication of a unitary matrix and a state-vector}
	\label{fig:qmdd_mult}
\end{figure*}

\subsection{Kronecker Product}
\label{sec:kronecker}

As discussed in Section~\ref{sec:simulation}, forming the Kronecker product of two matrices is essential to construct $2^n \times 2^n$  unitary matrices. 
Since we deal with square matrices whose dimensions are  powers of 2 when  simulating quantum computations, the Kronecker product is  defined as

\[A\otimes B = \left[ \begin{matrix}
a_{0,0}\cdot B & \cdots & a_{0,2^k-1}\cdot B \\
\vdots & \ddots & \vdots \\
a_{2^k-1,0}\cdot B & \cdots & a_{2^k-1,2^k-1} \cdot B
\end{matrix} \right].\]

This means that each element $a_{i,j}$ of $A$ has to be replaced by $a_{i,j}\cdot B$. While this constituted an expensive task using array-based realizations of $A$ and $B$, it is very cheap to form the Kronecker product of two matrices given in the proposed decision diagram. This has also already been observed in~\cite{niemann2017efficient}.

Since $a_{i,j}$ is given as product of the edge weights from~$A$'s root node to the terminal and we can easily determine $a_{i,j}\cdot B$ by adjusting the weight of the edge pointing to $B$'s root node. All that has to be done to determine $A\otimes B$ is replacing $A$'s terminal with the root node of $B$. Additionally, the weight of $A$'s root edge has to be multiplied by the weight of $B$'s root edge.

\begin{example}
	Recall the matrices considered in Fig~\ref{fig:dd_kronecker}. The Kronecker product $U = H \otimes I_2$ was efficiently constructed by taking the decision diagram representation of $H$ (shown in Fig.~\ref{fig:dd_h}) and replacing its terminal node with the root node of the decision diagram representing $I_2$ (shown in Fig.~\ref{fig:dd_i}). Since the root edge of $I_2$ has weight 1, the value of the root node of $U$ is equal to the weight of $A$'s root edge.
\end{example}

Note that forming the Kronecker product is not that simple when using other types of decision diagrams like QuIDDs. Here, the complex entries of the matrices are stored in different terminals rather than in edge weights. Consequently, forming the Kronecker product of two matrices $A$ and $B$ represented by QuIDDs requires -- among others -- to multiply all terminals of $A$ with those of $B$. 

\subsection{Multiplying Unitary Matrices}
\label{sec:dd_quantum_op}

The vector/matrix-multiplication as defined in~Def.~\ref{def:qua_op} can also be decomposed with respect to the most significant qubit leading to

\[
\psi'_i = \sum_{k=0}^{2^n-1}u_{i,k}\cdot \psi_k = \sum_{k=0}^{2^{n-1}-1}u_{i,k}\cdot \psi_k + \sum_{k=2^{n-1}}^{2^n-1}u_{i,k}\cdot \psi_k,
\]  

or, using the matrix notation, 
\[
	U \cdot \psi = \begin{bmatrix} U_{00} & U_{01} \\ U_{10} & U_{11} \end{bmatrix} \cdot \begin{bmatrix} \psi_0 \\ \psi_1 \end{bmatrix} = \begin{bmatrix} U_{00} \cdot \psi_0 \\ U_{10}\cdot \psi_0 \end{bmatrix} + \begin{bmatrix} U_{01} \cdot \psi_1 \\ U_{11}\cdot \psi_1 \end{bmatrix}.
\]
\medskip

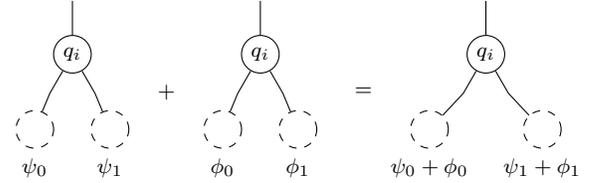
\begin{figure}
	\centering
	\begin{tikzpicture}[terminal/.style={draw,rectangle,inner sep=2pt}]
	
	\matrix[matrix of nodes, ampersand replacement=\&,every node/.style={vertex},column sep={1cm,between origins},row sep={1cm,between origins}] (qmdd3) {
		\& \node (top3)[draw = none] {};\& \\
		\& \node (o1) {$q_{i}$};\& \\
		\node[dashed, xshift=0.5cm] (o2) {$\phantom{q_i}$}; \& \node[dashed, xshift=0.5cm] (o3) {$\phantom{q_i}$}; \& \\
	};
	
	\draw (top3) -- (o1);
	
	\draw (o1) -- ++(240:0.6cm) -- (o2);
	\draw (o1) -- ++(300:0.6cm) -- (o3);
	
	\matrix[matrix of nodes, right=0.25cm of qmdd3, ampersand replacement=\&,every node/.style={vertex},column sep={1cm,between origins},row sep={1cm,between origins}] (qmdd4) {
		\& \node (top4)[draw = none] {};\& \\
		\& \node (l1) {$q_{i}$};\& \\
		\node[dashed, xshift=0.5cm] (l2) {$\phantom{q_i}$}; \& \node[dashed, xshift=0.5cm] (l3) {$\phantom{q_i}$}; \& \\
	};
	
	\draw (top4) -- (l1);
	
	\draw (l1) -- ++(240:0.6cm) -- (l2);
	\draw (l1) -- ++(300:0.6cm) -- (l3);
	
	\matrix[matrix of nodes, right=0.25 of qmdd4, ampersand replacement=\&,every node/.style={vertex},column sep={1.5cm,between origins},row sep={1cm,between origins}] (qmdd) {
		\& \node (top)[draw = none] {};\& \\
		\& \node (n1) {$q_{i}$};\& \\
		\node[dashed, xshift=0.75cm] (n2) {$\phantom{q_i}$}; \& \node[dashed, xshift=0.75cm] (n3) {$\phantom{q_i}$}; \& \\
	};
	
	\draw (top) -- (n1);
	
	\draw (n1) -- ++(240:0.6cm) -- (n2);
	\draw (n1) -- ++(300:0.6cm) -- (n3);
	
	\draw (o2.south) node[anchor=north] {$\psi_{0}$};
	\draw (o3.south) node[anchor=north] {$\psi_{1}$};
	
	\draw (l2.south) node[anchor=north] {$\phi_{0}$};
	\draw (l3.south) node[anchor=north] {$\phi_{1}$};
	
	\draw (n2.south) node[anchor=north] {$\psi_{0} + \phi_0$};
	\draw (n3.south) node[anchor=north] {$\psi_{1} + \phi_1$};				
	
	\draw ($(o3)!0.5!(l2) + (0,0.5)$) node {$+$};
	
	\draw ($(n2)!0.5!(l3) + (0,0.5)$) node {$=$};

	\end{tikzpicture}	
	
	\caption{Addition of state-vectors}
	\label{fig:qmdd_add}
\end{figure}

This means, that we have to recursively determine\footnote{The decompositions of multiplication and addition are recursively applied until $1\times 1$ matrices or 1-dimensional vectors result. Since those eventually represent just complex numbers, their multiplication and/or addition is straight-forward.} the four sub-products $U_{00} \cdot \psi_0$, $U_{01} \cdot \psi_1$, $U_{10} \cdot \psi_0$, and $U_{11} \cdot \psi_1$.
As shown in Fig.~\ref{fig:qmdd_mult}, these sub-products are then combined with a decision diagram node to two intermediate state vectors. 
Finally, these intermediate state vectors have to be added. This addition can be recursively decomposed in a similar fashion, namely

\[
\psi + \phi = \begin{bmatrix}\psi_0 \\\psi_1 \end{bmatrix} +  \begin{bmatrix}\phi_0 \\\phi_1 \end{bmatrix} = \begin{bmatrix}\psi_0 + \phi_0 \\ \psi_1 + \phi_1 \end{bmatrix}.
\]
The recursively determined sub-sums $\psi_0 + \phi_0$ and $\psi_1 + \phi_1$ are composed by a decision diagram node as shown in Fig.~\ref{fig:qmdd_add}.

Moreover, all these decompositions into sub-products and sub-sums do not change the decision diagram structure. \linebreak Hence, the complexity of them remains bounded by the number of nodes of the original representations. Furthermore, redundancies can again be exploited by caching sub-products and sub-sums. 

\subsection{Measurement}

Measurement can also be conducted efficiently on the decision diagram structure. 
To this end, consider w.l.o.g.~that qubit~$q_0$ (which is represented by the root node of the corresponding decision diagram) of the state vector should be measured (this can easily be accomplished by applying a SWAP operation or by re-arranging the nodes and edges of the decision diagram). 
Then, the left (right) successor of the root node represents the sub-vector containing the amplitudes of all states with $q_0=\ket{0}$ ($q_0=\ket{1}$), i.e.~states that are of form $\ket{0q_1q_2\ldots}$ ($\ket{1q_1q_2\ldots}$). The probability for collapsing qubit $q_0$ to one of the two basis states is defined as follows.

\begin{definition}
Consider a quantum system composed of $n$ qubits $q_0, q_1, \cdots q_{n-1}$. Then, the probability measuring (and, thus, collapsing to) basis state~$\ket{0}$ (basis state~$\ket{1}$) for qubit $q_0$ is the sum of the squared magnitudes of the complex entries in the corresponding sub-vector, i.e.
\begin{alignat*}{1}
P(q_0\rightarrow \ket{0}) &= \sum_{x \in 0\{0,1\}^{n-1}}\left|\alpha_x\right|^2 \\
P(q_0 \rightarrow \ket{1}) &= \sum_{x \in 1\{0,1\}^{n-1}}\left|\alpha_x\right|^2.
\end{alignat*}
	
\end{definition} 

\begin{example}
	Consider again the quantum state discussed in Example~\ref{ex:dd_vector}. The probabilities for measuring $q_0=\ket{0}$ and \mbox{$q_0=\ket{1}$} are:
	
	\begin{alignat*}{2}
	P(q_0\rightarrow \ket{0}) &= \left|0\right|^2 + \left|0\right|^2 + \left|\frac{1}{2}\right|^2 + \left|0\right|^2 &= \frac{1}{4} \\
	P(q_0\rightarrow \ket{1}) &= \left|\frac{1}{2}\right|^2 + \left|0\right|^2 + \left|-\frac{1}{\sqrt{2}}\right|^2 + \left|0\right|^2 &= \frac{3}{4} \\
	\end{alignat*}
	Therefore, the qubit $q_0$ is collapsed into basis state \ket{0} (basis state \ket{1}) with a probability of 0.25 (0.75). 
\end{example}

\begin{figure}
	\centering
	\begin{tikzpicture}[terminal/.style={draw,rectangle,inner sep=2pt}]

	\matrix[matrix of nodes, right=0.25 of qmdd4, ampersand replacement=\&,every node/.style={vertex},column sep={1cm,between origins},row sep={1cm,between origins}] (qmdd) {
		\& \node (n1) {$q_{i}$};\& \\
		\node[dashed, xshift=0.5cm] (n2) {$\phantom{q_i}$}; \& \node[dashed, xshift=0.5cm] (n3) {$\phantom{q_i}$}; \& \\
	};
	
	\draw (n1.north) node[anchor=south]{$p = p_{left} \cdot w_l^2 + p_{right} \cdot w_r^2$};
	\draw (n1) -- ++(240:0.6cm) node[left]{$w_{l}$} -- (n2);
	\draw (n1) -- ++(300:0.6cm) node[right]{$w_r$}-- (n3);
		
	\draw (n2.south) node[anchor=north] {$p_{left}$};
	\draw (n3.south) node[anchor=north] {$p_{right}$};				
	
	\end{tikzpicture}	
	
	\caption{Probability of a decision diagram node}
	\label{fig:measurement}
\end{figure}
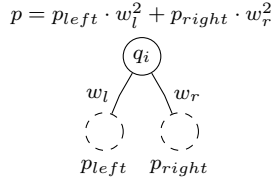

Consequently, we have to determine the summed probabilities for the decision diagram nodes. 
Again, the calculation of these probabilities can recursively be decomposed since 
\[
\sum_{x \in 0\{0,1\}^{n-1}}\left|\alpha_x\right|^2 = \sum_{x \in 00\{0,1\}^{n-2}}\left|\alpha_x\right|^2 + \sum_{x \in 01\{0,1\}^{n-2}}\left|\alpha_x\right|^2.
\]
This means we have to recursively determine the summed probabilities $p_{left}$ and $p_{right}$ of the sub-vectors. As Fig.~\ref{fig:measurement} shows, the summed probability of the current decision diagram node is then determined by the sum of the probabilities of the sub-vectors. Before these probabilities are added, they are multiplied with the squared weight of the respective edges.

	Note that, in the proposed decision diagram representation, the amplitudes $\alpha_x$ of the $2^n$ basis states are determined by a product of $n+1$ edge weights, i.e.~$\alpha_x = \prod_{i=0}^{n}w_{x,i}$. 
 Since $\left|\alpha \cdot \beta\right| = \left|\alpha\right| \cdot \left|\beta \right|$ holds for all complex numbers $\alpha,\beta \in \mathbb{C}$, $\left|\alpha_{x}\right|^2$ can be determined on the decision diagram, i.e.~
 \[
 	\left|\alpha_x\right|^2 = \left|\prod_{i=0}^{n}w_{x,i}\right|^2 = \prod_{i=0}^{n}\left|w_{x,i}\right|^2.
 \]

Finally, the weight on the edges to the left and the right successor of the root node (as well as the weight of the root edge) have to be considered to obtain the correct probabilities $P(q_0 \rightarrow \ket{0})$ and $P(q_0 \rightarrow \ket{1})$.
An example illustrates the idea.

\addtocounter{example}{-1}
\begin{example}
	The decision diagram shown in Fig~\ref{fig:dd_vector_b} represents the quantum state $\psi$ (cf.~Example~\ref{ex:dd_vector}). The probability of the node labeled $q_2$ can be determined by $1^2 + 0^2 = 1$. Based on that, the probabilities of the two nodes labeled $q_1$ can be determined. These are $0^2 \cdot 1 + 1^2 \cdot 1 = 1$ for the left node and $1^1 \cdot 1 + \left|-\sqrt{2}\right|^2 \cdot 1 = 3$ for the right node. From these nodes, we can determine the probabilities for collapsing $q_0$ to basis state \ket{0} or \ket{1} by
\[	\begin{aligned}	
		P(q_0 \rightarrow \ket{0}) &= \left(\frac{1}{2}\right)^2 \cdot 1^2 \cdot 1 &= \frac{1}{4} \\
		P(q_0 \rightarrow \ket{1}) &= \left(\frac{1}{2}\right)^2 \cdot 1^2 \cdot 3 &= \frac{3}{4}
	\end{aligned}~.\]
\end{example}

Having the probabilities for collapsing $q_0$ to basis state $\ket{0}$ and $\ket{1}$ allows to sample the new value for $q_0$. If we obtain basis state $\ket{0}$ (basis $\ket{1}$), the amplitudes for all basis states with $q_0=\ket{1}$ ($q_0=\ket{0}$) drop to zero. 
In the decision diagram, we perform this collapse by changing the right (left) outgoing edge of the root node to point to the terminal and attach weight zero. Finally, the remaining (\mbox{non-zero}) amplitudes in the state vector must be modified in order to fulfill the normalization constraint (cf.~Section~\ref{sec:qubits}). To this end, all amplitudes are divided by $\sqrt{P(q_0 \rightarrow \ket{0})}$ ($\sqrt{P(q_0 \rightarrow \ket{1})}$). This can easily be conducted on the decision diagram structure by modifying the weight of the root edge.

\begin{figure}
	\centering
	\begin{subfigure}[b]{0.2\textwidth}
		\centering
		\begin{tikzpicture}[terminal/.style={draw,rectangle,inner sep=0pt}]	
		\matrix[matrix of nodes,ampersand replacement=\&,every node/.style={vertex},column sep={1cm,between origins},row sep={1cm,between origins}] (qmdd2) {
			\node (m1) {$q_0$}; \\
			\node (m2b) {$q_1$}; \\
			\node (m3) {$q_2$}; \\
			\node[terminal] (t3) {1}; \\
		};
		
		\draw ($(m1)+(0,0.7cm)$) -- (m1) node[right, midway]{$\frac{1}{2}$};
		
		\draw (m1) -- ++(240:2.1cm) -- (t3) node[right, at start] {$0$};
		\draw (m1) -- ++(300:0.6cm) -- (m2b) node[right, at start] {$1$};
			
		\draw (m2b) -- ++(240:0.6cm) -- (m3) node[right, at start] {$1$};
		\draw (m2b) -- ++(300:0.6cm) -- (m3) node[right, midway] {$-\sqrt{2}$};
		
		\draw (m3) -- ++(240:0.6cm) -- (t3) node[right, at start] {$1$};
		\draw (m3) -- ++(300:0.6cm) -- (t3) node[right, at start] {$0$};
		
		\end{tikzpicture}
		\caption{Measure $q_0=\ket{1}$}
		\label{fig:dd_measurement_a}		
	\end{subfigure}
	\begin{subfigure}[b]{0.2\textwidth}
		\centering
		\begin{tikzpicture}[terminal/.style={draw,rectangle,inner sep=0pt}]	
		\matrix[matrix of nodes,ampersand replacement=\&,every node/.style={vertex},column sep={1cm,between origins},row sep={1cm,between origins}] (qmdd2) {
			\node (m1) {$q_0$}; \\
			\node (m2b) {$q_1$}; \\
			\node (m3) {$q_2$}; \\
			\node[terminal] (t3) {1}; \\
		};
		
		\draw ($(m1)+(0,0.7cm)$) -- (m1) node[right, midway]{$\frac{1}{\sqrt{3}}$};
		
		\draw (m1) -- ++(240:2.1cm) -- (t3) node[right, at start] {$0$};
		\draw (m1) -- ++(300:0.6cm) -- (m2b) node[right, at start] {$1$};
		
		\draw (m2b) -- ++(240:0.6cm) -- (m3) node[right, at start] {$1$};
		\draw (m2b) -- ++(300:0.6cm) -- (m3) node[right, midway] {$-\sqrt{2}$};
		
		\draw (m3) -- ++(240:0.6cm) -- (t3) node[right, at start] {$1$};
		\draw (m3) -- ++(300:0.6cm) -- (t3) node[right, at start] {$0$};
		
		\end{tikzpicture}
		\caption{Normalize amplitudes}
		\label{fig:dd_measurement_b}		
	\end{subfigure}
	\caption{Measurement of qubit $q_0$}
	\label{fig:dd_measurement}
\end{figure}
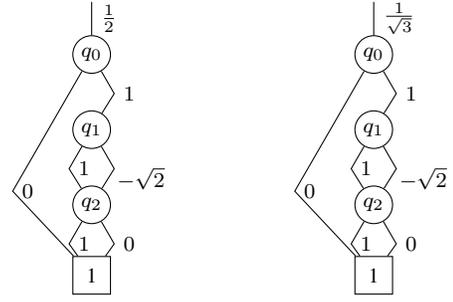

\addtocounter{example}{-1}
\begin{example}[continued]
	Assume we measure basis \linebreak state~$\ket{1}$ for qubit $q_0$. Fig.~\ref{fig:dd_measurement_a} shows the resulting decision diagram. To fulfill the normalization constraint, we divide the weight of the edge to the root node by $\sqrt{\frac{3}{4}}$ -- eventually resulting in the decision diagram shown in Fig.~\ref{fig:dd_measurement_b}.
\end{example}

Measuring all qubits can be conducted in a similar fashion. In fact, we repeat the procedure discussed above sequentially for all qubits $q_0, q_1, \cdots q_{n-1}$. Assume that qubit $q_i$ shall be measured, and that all qubits $q_{j}$ where $j<i$ are already measured. Then, there exists only one node labeled $q_i$, which is the root node of the sub-vector to be measured.

\section{Discussion}
\label{sec:discussion}

In this section, we discuss the complexity of the proposed approach with respect to existing array-based and \mbox{graph-based} solutions. 
This allows to get an intuition why the proposed approach can significantly outperform them in many cases.
We distinguish thereby between a discussion about the representation of vectors and matrices as well as a discussion about the respective operations.

\vspace*{-3mm}
\subsection{Representation of Vectors and Matrices} 

While array-based approaches always have to deal with \mbox{$2^n$-dimensional} state vectors and \mbox{$2^n\times 2^n$-dimensional} matrices, graph-based approaches often allow for a significantly more compact representation in many practically relevant cases. This is similar to BDDs in conventional design, which have an exponential size for most Boolean functions (especially random ones), but allow rather compact representation for many functions of interest. As already mentioned in Section~\ref{sec:vector}, introducing edge weights and a normalization scheme allows to exploit more redundancies than previous graph-based solutions -- leading to an even more compact representation. 

Nevertheless, in the worst case (i.e.~if no redundancies in the state vector can be exploited), a full binary tree with $\left|v\right| = 1 + \sum_{i=0}^{n-1} 2^i = 2^n$ nodes results. 
Furthermore, $2 \cdot (2^n-1) +1 = 2^{n+1} -1$ complex edge weights have to be stored -- approximately twice as many complex numbers than used in array-based solutions and when using e.g.~QuIDDs to represent the state-vector. That is, in the absolute worst case, the proposed representation is twice as large as \mbox{array-based} and graph-based solutions. This additional overhead, however, allows to exploit much more redundancies and, hence, to eventually gain more compact representations. This is confirmed by our experimental evaluation, which clearly shows that the peak node count of the proposed approach is significantly below that exponential upper bound.

In general, the worst-case memory complexity for $2^n \times 2^n$ matrices is analogous to the memory complexity for state-vectors -- a full quad-tree has $\left|v\right| = 1 + \sum_{i=0}^{n-1} 4^i = 1 + \frac{4^n-1}{3}$ nodes.
However, elementary quantum operations considered in this work (cf.~Section~\ref{sec:quantum_operations}) that are composed of a single target qubit and an arbitrary number of control qubits only require a linear (in the number of qubits) number of nodes (cf.~Section~\ref{sec:kronecker}). 

\vspace*{-3mm}
\subsection{Conducting Operations} Matrix operations are also less complex for graph-based solutions compared to array-based ones, which always suffer from an exponential complexity (since the exponentional matrix and the vector have to be traversed several times). However, there are also differences between the individual graph-based solutions. 

As discussed in Section~\ref{sec:kronecker}, the Kronecker product of two matrices can easily be  determined on the proposed type of decision diagrams by exchanging the terminal of one matrix with the root node of the other matrix. Consequently, forming the Kronecker product has complexity of $O(|v|)$. In contrast, this is more complex when using e.g.~QuIDDs, where it is required -- among others --  to multiply all terminal values of the two matrices with each other.

Also matrix vector multiplication can be performed significantly faster for graph-based solutions. The complexity of a matrix multiplication has an upper bound defined by the product of the number of nodes needed to represent the \mbox{state-vector} and the number of nodes needed to represent the matrix. Since we only consider matrices where the number of nodes grows linearly with the number of qubits, the overall complexity is $O(n\cdot\left| v\right|)$. 

Also measuring all qubits of a state vector has complexity~$O(|v|)$. This is, because each node has to be traversed only once. Afterwards, each qubit can be measured in $O(1)$, starting at the top of the decision diagram -- resulting in an overall complexity of $O(\left|v\right| + n)$.

\bigskip

Overall, this clearly shows that graph-based solutions offer more efficient representation and manipulation of state vectors in many cases. Although graph-based approaches suffer from overhead caused by the decision diagram structure (and additional complex numbers), these approaches can outperform array-based solutions by exploiting redundancies. 
Since the proposed graph-based approach exploits even more redundancies than e.g.~QuIDDPro, we also observe a significant performance improvement compared to this representative. 
This is confirmed by our empirical evaluation summarized in the next section. 

\section{Experimental Results}
\label{sec:exp}

We evaluated the scalability of the proposed approach and compared it to the state-of-the-art.  
To this end, we implemented the simulator in C++\footnote{The implementation is publicly available at \\\url{http://iic.jku.at/eda/research/quantum_simulation}} on top of the QMDD package provided by~\cite{DBLP:journals/tcad/NiemannWMTD16}, which we extended and modified to realize the concepts introduced above.
As state-of-the-art, we considered the publicly available implementations of the recently proposed array-based simulators~\emph{LIQ\textit{Ui}\ket{}}~\cite{DBLP:journals/corr/WeckerS14}, \emph{QX}~\cite{qxSimulator2017} and the simulator of \emph{ProjectQ}~\cite{steiger2016projectq}\footnote{Note that ProjectQ also provides an emulator, where high level operations are applied directly. However, in order to allow for a fair evaluation, we only compared the simulators against each other and not a simulator against an emulator.},
as well as the graph-based simulator \emph{QuIDDPro}~\cite{DBLP:books/daglib/0027785}. 
All simulations have been conducted on a regular Desktop computer, i.e.~a \mbox{64-bit} machine with 4 cores (8 threads) running at a clock frequency of 3.8~GHz and 32 GB of memory running Linux~4.4.\footnote{The proposed approach as well as QuIDDPro use a single core while the simulators LIQ\textit{Ui}\ket{}, QX, and the simulator of ProjectQ use multiple threads.}
Besides that, we additionally considered the best results published for other simulators (cf.~Section~\ref{sec:sota}) that have been taken from the respective papers. 

As benchmarks, well-known quantum algorithms considered by previous work have been used. More precisely, quantum systems generating entangled states, conducting \emph{Quantum Fourier Transformation}~(QFT; cf.~\cite{NC:2000}), executing \linebreak Grover's Algorithm for database search~\cite{DBLP:conf/stoc/Grover96}, and executing Shor's Factorization Algorithm~\cite{DBLP:journals/siamcomp/Shor97} (using the realization proposed by Beauregard~\cite{DBLP:journals/qic/Beauregard03} that requires $2n+3$ qubits to factor an $n$-bit integer) have been considered. 
Note that, for all benchmarks except QFT, the initial assignments of the inputs are fixed. For QFT, we randomly chose one of the basis states as initial input assignment.

In order to not run into numerical issues when normalizing the decision diagrams (which requires many divisions), we used the \emph{GNU MPFR} library~\cite{mpfr} to increase the precision of the floating point numbers and checked at each measurement whether the probabilities for measuring one of the basis states sum up to 1 (except a tiny $\epsilon$). In fact, using a precision of 200 bits was enough the avoid numerical errors for all considered benchmarks.

\begin{table*}[ht]
	\caption{Experimental results}
	\label{tab:results}
	\setlength{\tabcolsep}{2pt}
	\scriptsize
	\centering	
	\begin{tabular}{l|rr|rr|rr|rr|rrr|rrr}
		& & & \multicolumn{2}{c|}{LIQ\textit{Ui}\ket{}~\cite{DBLP:journals/corr/WeckerS14}} & \multicolumn{2}{c|}{QX~\cite{qxSimulator2017}} & \multicolumn{2}{c|}{ProjectQ~\cite{steiger2016projectq}} & \multicolumn{3}{c|}{QuIDDPro~\cite{DBLP:books/daglib/0027785}} & \multicolumn{3}{c}{Proposed approach} \\
		Computation & \#Qubits & \#Ops & Time\,[s] & Mem\,[MB] & Time\,[s] & Mem\,[MB] & Time\,[s] & Mem\,[MB] & Time\,[s] & Mem\,[MB]  & \#Nodes & Time\,[s] & Mem\,[MB]  & \#Nodes \\\hline\hline
		\begin{filecontents*}{csv/results_entanglement.csv}
Name,Qubits,Gates,Time,Mem,Time,Mem,Time,Mem,Time,Max Nodes,Mem,Time,Max Nodes,Mem
Entanglement,22,22,3.53,193.33,0.42,200.47,1.08,152.41,0.04,45,14.93,<0.01,43,48.02
,23,23,4.09,248.25,0.80,396.94,0.49,248.01,0.04,47,14.92,<0.01,45,48.03
,24,24,n.a.,,1.61,790.23,0.64,444.69,0.04,49,14.93,<0.01,47,48.00
,26,26,n.a.,,6.82,3149.66,2.74,1624.56,0.04,53,14.93,<0.01,51,48.02
,28,28,n.a.,,30.53,12586.87,4.91,6342.95,0.04,57,14.93,<0.01,55,48.14
,29,29,n.a.,,63.02,25169.79,9.12,12634.67,0.04,59,14.92,<0.01,57,48.24
,30,30,n.a.,,n.a.,,17.76,25217.66,0.04,61,14.93,<0.01,59,48.14
,31,31,n.a.,,n.a.,,MO,,0.04,63,14.93,<0.01,61,48.11
,100,100,n.a.,,n.a.,,MO,,0.14,201,15.98,<0.01,199,49.32
\end{filecontents*}
\csvreader[
		late after line=\\,
		late after last line=\\\cline{2-15},
		]{csv/results_entanglement.csv}
		{1=\Name,2=\QBW, 3=\Ngates, 4=\TimeLi, 5=\MemLi, 6=\TimeQX, 7=\MemQX, 8=\TimePQ, 9=\MemPQ, 10=\TimeQP, 12=\MemQP, 11=\MaxNodesQP, 13=\Time, 15=\Mem, 14=\MaxNodes}
		{\Name  & \QBW & \Ngates & \TimeLi & \MemLi & \TimeQX & \MemQX & \TimePQ & \MemPQ & \TimeQP & \MemQP & \MaxNodesQP & \Time & \Mem & \MaxNodes}
		&\multicolumn{9}{l}{~~\textbullet~Reported for~QX~\cite{qxSimulator2017}: max. 34 qubits using less than 270 GB of memory} \\\hline\hline
		\begin{filecontents*}{csv/results_qft.csv}
Name,Qubits,Gates,Time,Mem,Time,Mem,Time,Mem,Time,Mem,Nodes,Time,Max Nodes,Mem,Comp
QFT,18,171,3.02,192.91,0.27,16.56,0.87,57.82,24.21,65535,192.77,0.01,18,48.47,
,20,210,4.35,188.67,1.59,53.49,0.50,75.82,2263.38,2097151,1210.34,0.01,20,48.56,
,21,231,6.46,192.83,3.63,102.75,0.66,100.75,10208.06,4194303,2511.26,0.01,21,48.78,
,22,253,11.38,191.00,8.06,201.01,1.06,150.30,MO,,,0.01,22,48.92,
,23,276,22.43,312.41,15.06,397.50,1.55,250.18,MO,,,0.01,23,49.13,
,24,300,n.a.,,31.66,790.77,3.27,445.04,MO,,,0.01,24,49.19,
,26,351,n.a.,,139.23,3150.14,12.84,1624.81,MO,,,0.02,26,49.36,
,29,435,n.a.,,1270.03,25170.21,109.19,12638.54,MO,,,0.02,29,49.95,
,30,465,n.a.,,n.a.,25217.39,234.27,,MO,,,0.02,30,50.00,
,31,496,n.a.,,n.a.,,MO,,MO,,,0.03,31,50.53,
,64,2080,n.a.,,n.a.,,MO,,MO,,,0.09,64,68.00,
\end{filecontents*}
\csvreader[ 
		late after line=\\,
		late after last line=\\\cline{2-15},
		]{csv/results_qft.csv}
		{1=\Name,2=\QBW, 3=\Ngates, 4=\TimeLi, 5=\MemLi, 6=\TimeQX, 7=\MemQX, 8=\TimePQ, 9=\MemPQ, 10=\TimeQP, 12=\MemQP, 11=\MaxNodesQP, 13=\Time, 15=\Mem, 14=\MaxNodes}
		{\Name  & \QBW & \Ngates & \TimeLi & \MemLi & \TimeQX & \MemQX & \TimePQ & \MemPQ & \TimeQP & \MemQP & \MaxNodesQP & \Time & \Mem & \MaxNodes}	&	\multicolumn{9}{l}{~~\textbullet~Reported for qHiPSTER (Intel, cf.~\cite{DBLP:journals/corr/SmelyanskiySA16}): max. 40 qubits on a supercomputer (in approx. 1000\,s).} \\
		&\multicolumn{9}{l}{~~\textbullet~Reported for Quantum emulator from~\cite{DBLP:conf/sc/HanerSST16}: max. 36 qubits on a supercomputer (in approx. 10\,s).} \\\hline\hline
		\begin{filecontents*}{csv/results_grover.csv}
Name,Qubits,Gates,Time,Mem,Time,Mem,Time,Mem,Time,Max Nodes,Mem,Time,Max Nodes,Mem
Grover,16,11600,97.78,195.11,55.90,57.28,6.65,51.88,6.93,39,16.29,0.14,130,50.54
,18,26082,770.26,193.42,583.45,144.41,16.37,58.12,23.49,44,17.24,0.33,148,50.78
,20,57940,8494.59,198.22,6394.54,382.51,77.95,77.17,85.86,39,19.05,0.78,166,50.80
,21,86037,>18000.00,,>18000.00,,229.45,101.27,168.07,n.a.,20.60,0.97,175,50.91
,24,278040,n.a.,,>18000.00,,5362.00,447.11,1272.36,n.a.,31.78,3.11,202,51.14
,25,409625,n.a.,,>18000.00,,15765.00,843.85,2598.08,n.a.,40.55,5.55,211,51.14
,26,602394,n.a.,,>18000.00,,>18000.00,,5076.64,n.a.,52.95,8.24,220,51.26
,27,884763,n.a.,,>18000.00,,>18000.00,,>18000.00,,,14.65,238,51.26
,30,2780430,n.a.,,n.a.,,>18000.00,,>18000.00,,,37.23,256,51.32
,40,118632840,n.a.,,n.a.,,>18000.00,,>18000.00,,,1239.96,346,52.01
\end{filecontents*}
\csvreader[ 
		late after line=\\,
		late after last line=\\\hline,
		]{csv/results_grover.csv}
		{1=\Name,2=\QBW, 3=\Ngates, 4=\TimeLi, 5=\MemLi, 6=\TimeQX, 7=\MemQX, 8=\TimePQ, 9=\MemPQ, 10=\TimeQP, 12=\MemQP, 11=\MaxNodesQP, 13=\Time, 15=\Mem, 14=\MaxNodes}
		{\Name  & \QBW & \Ngates & \TimeLi & \MemLi & \TimeQX & \MemQX & \TimePQ & \MemPQ & \TimeQP & \MemQP & \MaxNodesQP & \Time & \Mem & \MaxNodes}		\begin{filecontents*}{csv/results_shor.csv}
Name,Qubits,Gates,Time,Mem,Time,Mem,Time,Mem,Time,Max Nodes,Mem,Time,Max Nodes,Mem,Comp
Shor,13,12640,76.75,65.88,n.a.,,0.40,47.56,1665.28,16375,92.03,0.21,40,52.65,
,15,23083,298.59,62.72,n.a.,,9.30,51.43,16236.14,65535,365.22,0.54,72,55.09,
,17,38847,343.83,128.81,n.a.,,19.25,55.10,>18000.00,,,0.76,66,57.63,
,19,61510,1232.83,85.01,n.a.,,47.48,70.91,>18000.00,,,1.07,156,60.56,
,21,92700,7888.00,147.03,n.a.,,187.14,115.74,>18000.00,,,10.98,519,64.88,
,23,134490,>18000.00,,n.a.,,947.72,283.88,>18000.00,,,3.35,151,67.30,
,25,188784,n.a.,,n.a.,,4827.00,860.96,>18000.00,,,17.91,653,75.17,
,27,258060,n.a.,,n.a.,,>18000.00,,>18000.00,,,74.52,949,83.34,
,31,451577,n.a.,,n.a.,,MO,,>18000.00,,,44.60,305,97.91,
,33,581272,n.a.,,n.a.,,MO,,>18000.00,,,1019.55,6517,156.12,
,37,922385,n.a.,,n.a.,,MO,,>18000.00,,,5585.64,20917,259.96,
\end{filecontents*}
\csvreader[ 
		late after line=\\,
		late after last line=\\\cline{2-15},
		]{csv/results_shor.csv}
		{1=\Name,2=\QBW, 3=\Ngates, 4=\TimeLi, 5=\MemLi, 6=\TimeQX, 7=\MemQX, 8=\TimePQ, 9=\MemPQ, 10=\TimeQP, 12=\MemQP, 11=\MaxNodesQP, 13=\Time, 15=\Mem, 14=\MaxNodes}
		{\Name  & \QBW & \Ngates & \TimeLi & \MemLi & \TimeQX & \MemQX & \TimePQ & \MemPQ & \TimeQP & \MemQP & \MaxNodesQP & \Time & \Mem & \MaxNodes}
		&\multicolumn{9}{l}{~~\textbullet~Reported for LIQ\textit{Ui}\ket{} (Microsoft, cf.~\cite{DBLP:journals/corr/WeckerS14}): max.~31 qubits (in more than 30 days).} \\\hline
		\end{tabular}
		\\
		\raggedright{Time denotes the actual runtime and not the CPU seconds (which would be even higher for~\cite{DBLP:journals/corr/WeckerS14,qxSimulator2017,steiger2016projectq} since they use parallelization to speed-up simulation).}
\end{table*}

Table~\ref{tab:results} summarizes the results.
The columns \emph{\#Qubits} and \emph{\#Ops} 
list the number of involved qubits and the number of quantum operations to be conducted, respectively. 
In the remaining columns, we list the simulation time (i.e.~the entire runtime from initialization to termination) and peak memory when using LIQ\textit{Ui}\ket{}, QX, ProjectQ, QuIDDPro, as well as the proposed approach. 
For the \mbox{graph-based} simulators (i.e.~QuIDDPro as well as the proposed one), we also list the peak node count during simulation, which gives a more accurate measurement of scalability than the actual memory consumption.  
Besides that, Table~\ref{tab:results}  also provides a comparison to other state-of-the-art simulation approaches. That is, whenever results from them are provided in the literature,  
the respectively best result for a considered quantum algorithm is summarized in the bottom of that part.

Note that some restrictions apply for certain \mbox{state-of-the-art} simulators:
In fact, the publicly available version of LIQ\textit{Ui}\ket{} allows to simulate circuits composed of at most 23 qubits only. 
Using QX, we were able to simulate up to 29 qubits on our machine -- trying to allocate 30 or more qubits failed (due to limited memory).
Furthermore, QX does not allow to simulate Beauregard's realization of Shor's algorithm for integer factorization, because of missing features in the circuit description language (since QX is still in its infancy). We have accordingly marked all these cases 
by \emph{n.a.} (not applicable) in Table~\ref{tab:results}. 
Furthermore, the current release of QuIDDPro (version~3.8) also contains an improved simulator called \emph{QuIDDProLite} (to be activated with the command line option~\mbox{\emph{-cs}}) that runs on average four times faster than QuIDDPro. However, this improved version can only simulate stand-alone quantum circuits and, hence, is not applicable for simulating Beauregard's implementation of Shor's algorithm (which also requires non-quantum control structures). Consequently, we list the runtime of QuIDDPro for Shor's algorithm and the runtime of QuIDDProLite for the other benchmarks (i.e.~always the best result of both QuIDDPro versions are reported).
Since QuIDDProLite does not offer the capability to dump the peak node count, we list the number of nodes obtained when using QuIDDPro for simulation. In the cases where this simulation does not succeed within the given timeout of five hours, we again label the corresponding entry by \emph{n.a.} (not applicable).

As can be seen,  
the simulation of quantum systems generating entangled states and conducting QFT
shows a linear behavior on our simulator.
While this allows for a rather unlimited
scalability using the solution proposed in this work, Microsoft's simulator LIQ\textit{Ui}\ket{}, QX, as well as ProjectQ 
show exponential behavior.
Even massive hardware power such as employed by Intel's simulator \emph{qHiPSTER}~\cite{DBLP:journals/corr/SmelyanskiySA16} (running on 
a machine with 1000 nodes and 32\,terabytes of memory) 
or the quantum emulator of~\cite{DBLP:conf/sc/HanerSST16} (running on a similar machine)
manages to conduct QFT for a maximum of 40 qubits only (and additionally requires hundreds of seconds, while the approach proposed in this work terminates in a fraction of a second). 

The graph-based simulator \mbox{QuIDDPro~\cite{DBLP:books/daglib/0027785}} is capable to efficiently conduct entanglements (similar to the proposed approach, only a  linear amount of nodes is required). But also here, an exponential behavior can be observed when conducting QFT. This is caused by the fact that the state vector contains exponentially many different entries which cannot be handled efficiently by QuIDDPro. In contrast, the proposed solution can exploit further redundancies here  (namely sub-vectors which are multiples of each other as discussed in Section~\ref{sec:representation}) -- resulting in a linear number of decision diagram nodes for QFT.

The simulation of Grover's Algorithm and Shor's Algorithm constitutes a more challenging task. 
But even here, the proposed representation remains rather compact.
For example, in case of simulating Shor's Algorithm with 37 qubits, only slightly more than 20\,000
nodes are required. 
In fact, the significantly larger number of operations 
is more challenging here. Nevertheless, the proposed approach still manages to simulate both algorithms significantly more efficient and for
more qubits than the \mbox{state-of-the-art}. While e.g.~Microsoft's simulator \emph{LIQ\textit{Ui}\ket{}} 
is capable of conducting Shor's Algorithm
for at most 31 qubits in more than 30 days (on a similar machine; cf.~\cite{DBLP:journals/corr/WeckerS14}), the simulation approach proposed in this work completes this task within in less than a minute.

Also the previously proposed graph-based solution \mbox{QuIDDPro} can not reach this efficiency. While QuIDDPro  allows for a rather compact representation when simulating Grover's Algorithm, it requires a substantial amount of runtime. This is caused by the fact that the required operations cannot be conducted as efficiently as with the proposed solution (since the decomposition scheme is not that natural to matrices and state-vectors as discussed in Section~\ref{sec:discussion}). Hence,  the proposed solution clearly outperforms QuIDDPro (which is optimized for Grover's Algorithm and has mainly been evaluated on that in the literature). For Shor's Algorithm, we can observe similar limitations for  QuIDDPro than for array-based solutions (in fact, LIQ\textit{Ui}\ket{} is performing even better than QuIDDPro in this case). Again, the solution proposed in this work can handle all these cases much faster and for more qubits than before.

Overall, the proposed simulation approach clearly outperforms the current state-of-the-art in terms of runtime as well as in terms of memory (only up to 260 MB were required).
Additionally, the proposed solution is able to simulate quantum computations for more qubits. Besides that, all these accomplishments can be achieved on a single core of a regular Desktop machine, i.e.~without massive hardware power or the utilization of supercomputers. 

\section{Conclusions}
\label{sec:conclusion}

This work introduces a new graph-based approach for the simulation of quantum computations that clearly outperform previous graph-based or array-based solutions.
To this end, we revisited the basics of quantum computation and developed a simulation approach which exploits redundancies in the respective quantum state and operation descriptions. The resulting simulator (which is publicly available at \mbox{\url{http://iic.jku.at/eda/research/quantum_simulation}}) is capable of 
(1)~simulating quantum computations for more qubits than before, 
(2)~in significantly less run-time (in hours or, in many cases, just minutes or seconds rather than several days), and  
(3)~on a regular Desktop machine. 

\section*{Acknowledgements}
The authors would like to thank Igor Markov for many fruitful discussions. Furthermore, this works has partially been supported by the European Union through the COST Action IC1405 and the Google Research Award Program.

\newpage
\bibliographystyle{abbrv}
{
	\bibliography{bib/lit_header,bib/lit_mymisc,bib/lit_myrev,bib/lit_others,bib/lit_othersrev,bib/lit_rev,bib/lit_testing,bib/lit_misc,bib/lit_new,bib/lit_simulation,bib/lit_quantum}
}

\begin{IEEEbiography}
	[{\includegraphics[width=1in,height=1.25in,clip,keepaspectratio]{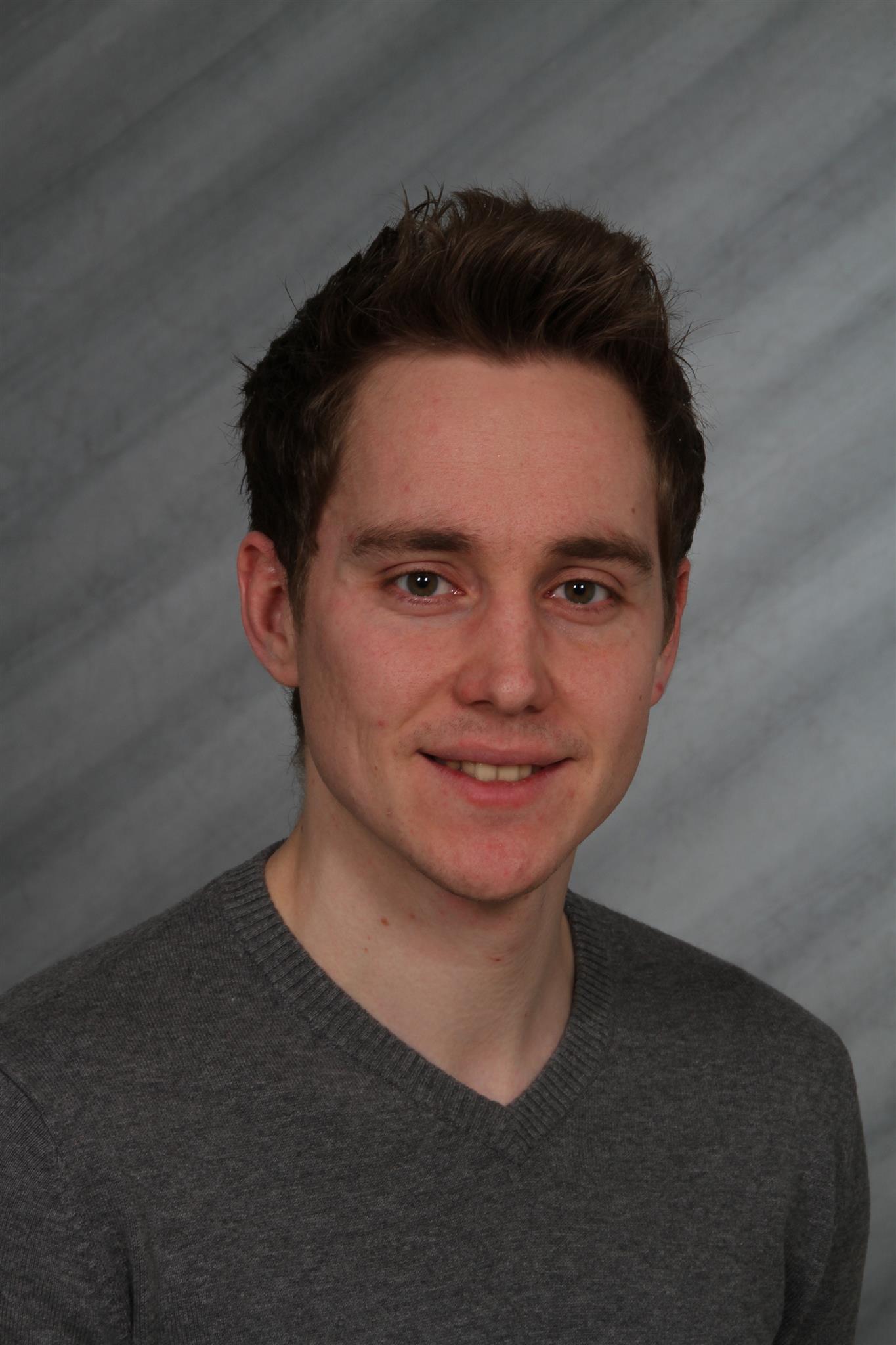}}]{Alwin Zulehner}
	Alwin Zulehner (S’17) received his BSc and MSc degree in computer science from the Johannes Kepler University Linz, Austria in 2012 and 2015, respectively.
	He is currently a Ph.D. student at the Institute for Integrated Circuits at the Johannes Kepler University Linz, Austria. 
	His research interests include design automation for emerging technologies, currently focusing on reversible circuits and quantum circuits. In these areas, he has published several papers on international conferences and journals such as the IEEE Transactions on Computer Aided Design of Integrated Circuits and Systems (TCAD), Asia and South Pacific Design Automation Conference (ASP-DAC), Design, Automation and Test in Europe (DATE) and International Conference on Reversible Computation.
\end{IEEEbiography}

\begin{IEEEbiography}
	[{\includegraphics[width=1in,height=1.25in,clip,keepaspectratio]{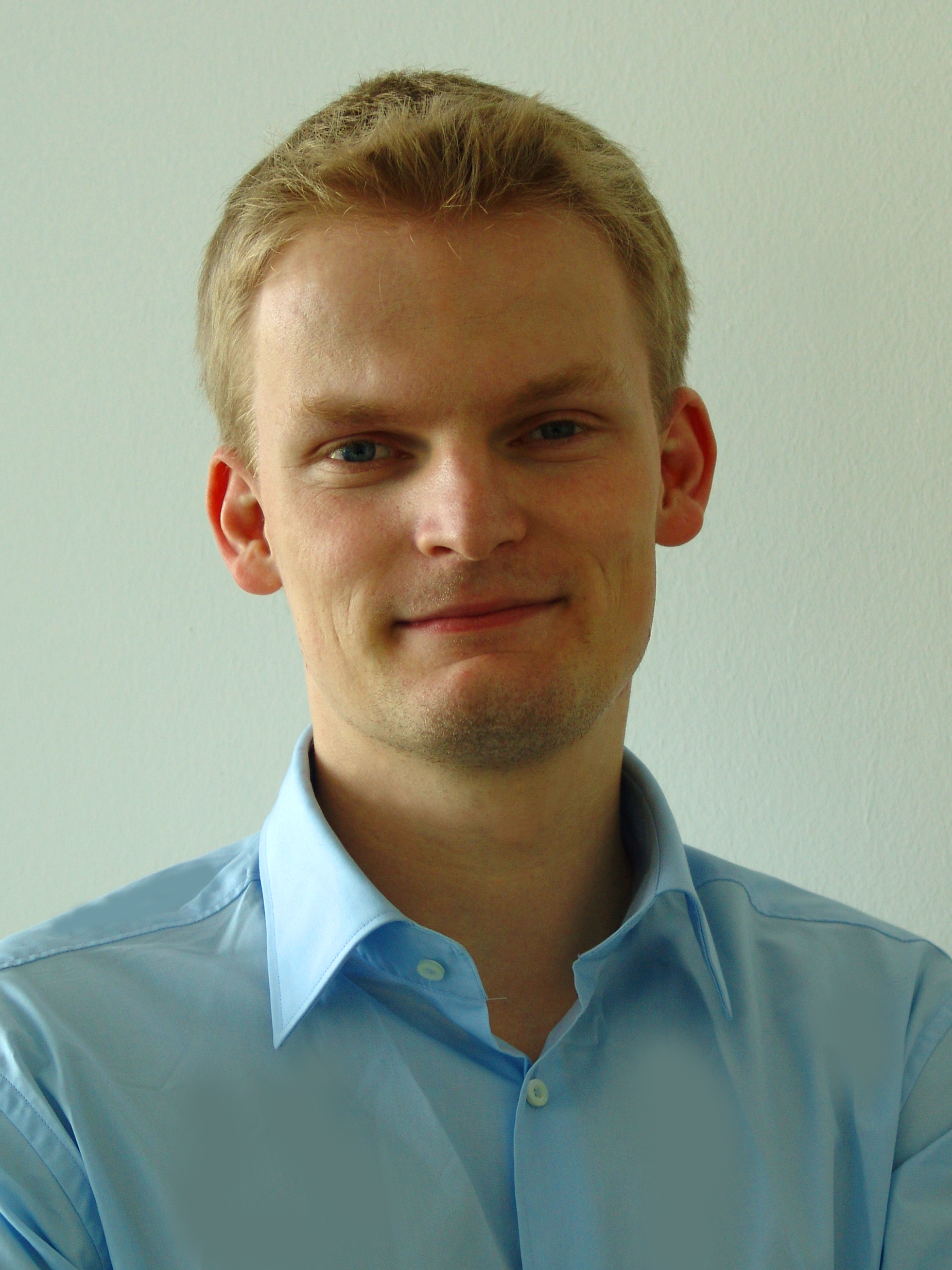}}]{Robert Wille}
Robert Wille (M’06–SM’09) received the Diploma and Dr.-Ing. degrees in computer science from the University of Bremen, Bremen, Germany, in 2006 and 2009, respectively. He was with the Group of Computer Architecture, University of Bremen, from 2006 to 2015, and has been with the German Research Center for Artificial Intelligence (DFKI), Bremen, since 2013. He was a Lecturer with the University of Applied Science of Bremen, Bremen, Germany, and a Visiting Professor with the University of Potsdam, Potsdam, Germany, and Technical University Dresden, Dresden, Germany. Since 2015, he is a Full Professor with Johannes Kepler University Linz, Linz, Austria. His current research interests include the design of circuits and systems for both conventional and emerging technologies. In these areas, he has published over 200 papers in journals and conferences. Dr. Wille has served in Editorial Boards and Program Committees of numerous journals/conferences, such as the IEEE Transactions on Computer Aided Design of Integrated Circuits and Systems  (TCAD), Asia and South Pacific Design Automation Conference (ASP-DAC), Design Automation Conference (DAC), Design, Automation and Test in Europe (DATE) and International Conference on Computer Aided Design (ICCAD).
\end{IEEEbiography}

\flushend

\end{document}